\documentclass[11pt]{article}

\usepackage{color,amsthm,amsmath,natbib,algorithm,algpseudocode}
\usepackage{parskip}
\usepackage{amsfonts}
\usepackage{enumerate}
\usepackage{graphicx}
\usepackage{rotating}
\usepackage{natbib}
\usepackage{hyperref}

\newtheorem{prop}{Proposition}[section]

\newtheorem{theorem}{Theorem}[section]

\newcommand{\R}{\mathbb{R}}
\newcommand{\E}{\mathbb{E}}

\newcommand{\argmin}{\mathop{\arg \min}\limits}
\newcommand{\argmax}{\mathop{\arg \max}\limits}

\newcommand{\calN}{\mathcal{N}}
\newcommand{\Z}{\mathbf{Z}}
\newcommand{\D}{\mathbf{D}}

\newcommand{\calI}{\mathcal{I}}
\newcommand{\X}{\mathbf{X}}
\newcommand{\Y}{\mathbf{Y}}
\newcommand{\U}{\mathbf{U}}
\newcommand{\V}{\mathbf{V}}

\newcommand{\minI}{\mathring{I}}

\newcommand{\ID}{\mathcal{I}}

\DeclareMathOperator{\Var}{Var}

\begin{document}
\title{Two-Sample Testing in High-Dimensional Models}

\author{Nicolas St\"adler\\
\small Netherlands Cancer Institute\\[-0.8ex]
\small \small  Amsterdam, Netherlands.\\
\small \texttt{n.stadler@nki.nl}\\
\and
Sach Mukherjee\\
\small Netherlands Cancer Institute\\[-0.8ex]
\small \small  Amsterdam, Netherlands.\\
\small \texttt{s.mukherjee@nki.nl}\\
}


\date{}
\maketitle

\begin{abstract}
{We propose novel methodology for testing equality of model
  parameters between two high-dimensional populations. The technique 
  is very general and applicable to a wide range of models. The 
  method is based on sample splitting: the data is split into two
  parts; on the first part we reduce the dimensionality of the model
 to a manageable size; on the
  second part we perform significance testing (p-value calculation)
  based on a restricted likelihood ratio statistic. Assuming that both
  populations arise from the same distribution, we show that the
  restricted likelihood ratio statistic is asymptotically distributed
  as a weighted sum of chi-squares with weights which can be
  efficiently estimated from the data. In high-dimensional problems, a single data split can result in a ``p-value lottery". To ameliorate this effect, we iterate the splitting process and 
  aggregate the resulting p-values. This multi-split approach provides
  improved p-values. 
  We illustrate the use of our general 
  approach in two-sample comparisons of high-dimensional regression models (``differential regression") and 
  graphical models (``differential network"). In both cases we show results on simulated data as well as real data from recent, high-throughput cancer studies.}

\vspace{0.5cm}
\end{abstract}

\newpage
\section{Introduction and Motivation}



We consider the general two-sample testing problem where
the goal is to test whether or not two independent populations
$U$ and $V$,
parameterized by $\phi_u$ and $\phi_v$ respectively, with
$\phi_u,\phi_v \in \Phi\subset\R^p$, arise from the same distribution. The hypothesis testing problem of interest is
\begin{eqnarray}\label{eq:h0} 
\mathbf{H_0}:\phi_u=\phi_v \quad \textrm{against}\quad  \mathbf{H_A}:\phi_u\neq\phi_v.
\end{eqnarray}
In this paper the focus is on the high-dimensional setting where the number of
available samples per population ($n_u$ and $n_v$) is small compared to
the dimensionality of the parameter space~$p$.  

In a setup where $n_u$ and $n_v$ are much
larger than $p$ the
ordinary likelihood-ratio test offers a very general solution to problem~(\ref{eq:h0}). It is well-known that under $\mathbf{H_0}$ the likelihood ratio statistic is asymptotically $\chi^2_p$ distributed which
then allows computation of confidence intervals and p-values. However, in
settings where $p$ is large compared to sample sizes, the
likelihood-ratio statistic is ill-behaved and the classical asymptotic set-up cannot be relied upon to test statistical significance.

Our approach for solving the general high-dimensional two-sample
problem (\ref{eq:h0}) is motivated by the \emph{screen and clean}
procedure \citep{wasserman2009} 
originally developed for
improved variable selection in the high-dimensional regression
model. The idea is to split the data from both populations into
two parts; perform dimensionality reduction on one part and pursue
significance testing on the other part of the data. In more detail, on the first split we select three subsets of $\Phi$
by screening for the most relevant parameters of population $U$ and $V$
individually but also by selecting the important parameters of the pooled data from both populations. In this way we obtain an \emph{individual model}
with different parameter spaces for $U$ and $V$ and a \emph{joint model} which describes both populations by a common parameter
space. On the second split of the data, we then can compute the likelihood ratio statistic between these two models, the so-called \emph{restricted likelihood ratio statistic}. A crucial
observation is that the two models are
non-nested and therefore non-standard tools are needed to obtain the
asymptotic null distribution. We apply the theory on model selection and non-nested hypotheses 
developed by \cite{vuong1989} to the two-sample scenario and show that
the null distribution asymptotically approaches a weighted sum of
independent chi-squared random variables with weights which can be
efficiently estimated from the second split of the
data. Importantly, the weighted sum of chi-squared approximation is
invoked only in the second split, where the model dimensionalty has been reduced.

As indicated above our approach involves a screening or model
selection step prior to significance testing. Our method requires that
the parameter set selected by a specific screening procedure contains
the true model parameter (\emph{screening property}) and that the
selected set is small compared to the sample size (\emph{sparsity
  property}). The first property is needed for deriving the asymptotic
null distribution, whereas the latter property justifies
asymptotic approximation.  The two conditions which we impose here are much weaker than requiring consistency in variable
selection. We propose to use $\ell_1$-penalized maximum likelihood
estimation \citep{tibshirani96regression,fan01variable} with regularization
parameter chosen by cross-validation. This sets automatically
non-relevant parameter components to zero and allows for efficient
screening even in scenarios with very large $p$. $\ell_1$ penalization leads to sparse estimates and
there is theoretical evidence that the screening property 
 holds under mild additional assumptions.

In current applied statistics, a wide range of applications are faced with
high-dimensional data. Parameter estimation in the ``large $p$, small $n$''
setting has been extensively studied in theory \citep{lassobook2011} and also applied with
success in many areas. Since interpretation of parameters is crucial in applied science there is a clear 
need to assess uncertainty and statistical significance in such settings. However, significance testing in high-dimensional
settings has only recently attracted attention.  \cite{meinshausenpval2009} and \cite{buhlmann2012}
consider testing in the high-dimensional linear model. In
the context of high-dimensional two-sample comparison \cite{bai1996},
\cite{chen2010}, \cite{lopes2012} and others treat testing for differences in
the population means. And recently, \cite{tonycai2011} and \cite{li2012} developed a
two-sample test for covariance matrices of two high-dimensional populations. 

The approach proposed in this paper tackles the high-dimensional
two-sample testing in a very general setting. In our methodology $\phi_u$ and $\phi_v$ can parameterize any model of
interest. In the empirical examples we show below we focus on two specific applications of our approach:
(i) High-dimensional regression, where two populations may differ with
respect to regression models and (ii) Graphical modelling, where the
two populations may differ with respect to conditional independence structure.
In analogy to the term ``differential expression" as widely-used for testing means in gene expression studies, we call these ``differential regression" and ``differential network" respectively. Both high-dimensional regression and graphical models are now widely used in biological applications, and very often 
scientific interest focuses on potential differences between
populations (such as disease types, cell types, environmental
conditions etc.). However, to date in the high-dimensional setting, two sample testing concerning such models has not been well studied. 
The methodology we propose offers a way to directly test hypotheses concerning differences in molecular influences or biological network structure using high-throughput data.

The organization of the paper is as follows: Section~\ref{sec:method}
introduces the setup and the methodology: Section~\ref{sec:screen}
explains variable screening using $\ell_1$-regularization;
Section~\ref{sec:restricted.loglik} introduces the restricted
likelihood-ratio statistic; Sections~\ref{sec:asymptotic} and \ref{sec:weights} derives its
asymptotic null distribution and Section~\ref{sec:multisplit.pval}
shows how to compute
p-values using the single- and  multi-split algorithms. In
Section~\ref{sec:regr.ggm} we present and give details on the two specific examples of our
approach that we outlined above (\emph{differential regression} and \emph{differential
  network}). Finally, in Section~\ref{sec:exp.num}, we evaluate our
methodology on simulated and real data from two recent high-throughput studies in cancer biology.




\section{Data Splitting, Screening and Non-Nested Hypothesis}\label{sec:method}
Consider a conditional model class given by densities 
\begin{eqnarray}\label{eq:cond.model}
d(y|x;\phi),\;y\in\R^k,\;x\in\R^{l},\;z=(y,x)\;\textrm{and}\;
\phi\in\Phi\subset\R^p.
\end{eqnarray}
Note that $x$ can be empty (i.e., $l=0$). 
Let population $U=(Y_u,X_u)$ and $V=(Y_v,X_v)$, where $X_u$, $X_v$ are both
generated from the same distribution and conditional on the $X$'s,
$Y_u$ and $Y_v$ are generated from
$d(\cdot|x;\phi_u)$ and $d(\cdot|x;\phi_v)$ respectively.

The goal is to solve the general two-sample testing problem~(\ref{eq:h0})
given data-matrices $\U=(\Y_u,\X_u)\in\R^{n_u\times (k+l)}$ and $\V=(\Y_v,\X_v)\in\R^{n_v\times
  (k+l)}$, representing i.i.d. random samples of $U$ and $V$. We
consider the high-dimensional case with $p\gg \min\{n_v,n_u\}$ and assume that the data
generating parameters $\phi_u$ and $\phi_v$ are sparse, meaning
that many of their components are equal to zero. If $k=1,l=0$, we have a classical (univariate) two-sample set-up.
If $k=1$ and
$d(y|x;\phi)$ is the univariate Normal distribution with mean
$\beta^Tx$ and noise variance $\sigma^2$, $\phi=(\beta,\sigma^2)$,
then $U$ and $V$ follow linear regression models. If $l=0$ and
$d(y;\phi)$ is the multivariate Normal distribution with $\phi$
representing the inverse covariance matrix, then $U$ and $V$ follow a
Gaussian graphical model and (\ref{eq:h0}) asks whether or not two
graphical models (in short: networks) are significantly different. We
will treat these two examples in detail in Section~\ref{sec:regr.ggm}
and \ref{sec:exp.num}. 
We refer to the regression case as \emph{differential regression} and to the  graphical model case as \emph{differential network}. 

Our methodology for solving (\ref{eq:h0}) is based on sample splitting
and has its inspiration from the work by
\cite{wasserman2009}. We randomly divide data from population $U$ into two parts $\U_{\rm in}$ and $\U_{\rm out}$ of equal size and proceed in the same way with population $V$ which yields $\V_{\rm in}$ and $\V_{\rm out}$. 
In a first step the dimensionality of the parameter space $\Phi$ is
reduced by filtering out (potentially many) redundant components. We do this by
applying a screening procedure $\calI$ on $\U_{\rm in}$ and $\V_{\rm
  in}$ separately, but also on pooled data $(\U_{\rm in},\V_{\rm
  in})$. In this way we obtain models with lower dimensionality than the full dimension $p$. The first model describes populations $U$ and $V$
individually using different reduced-parameter spaces, whereas the second
model explains both populations jointly with a single reduced-parameter
space. In a second step, we then evaluate the \emph{restricted likelihood-ratio} statistic on the held-out data $\U_{\rm out}$ and $\V_{\rm out}$ and perform significance testing.

\subsection{Screening and $\ell_1$-regularization}\label{sec:screen}
Consider i.i.d. data $\Z=(\Y,\X)\in \R^{n\times (k+l)}$
with $Y|X=x$ distributed
according to $d(\cdot|x;\phi_o)$ and $\phi_o\in
\Phi\subset\R^{p}$. We have $p\gg n$ and $\phi_o$ is
supposed to be sparse.

A screening procedure $\calI$ selects, based on data $\Z$, a set $A$ of
active parameter components by a map $\calI(\Z) \subset \{1,\ldots,p\}$.
The map $\calI$ then defines the active parameter space
by $$\Phi_{\calI(\Z)}=\{(\phi_1,\ldots,\phi_p):\phi_j=0\;\textrm{for all}\;
j{\not\in}\calI(\Z)\}.$$
There are two basic requirements on the screening procedure
$\calI$. Firstly, the procedure should get rid of
many non-relevant components in the sense that the set of active
parameter components $\calI(\Z)$ should be small compared to $n$. Secondly, the active parameter space $\Phi_{\calI(\Z)}$
should contain the true parameter $\phi_o$. 
We refer to these
requirements as the \emph{sparsity} and \emph{screening} property:
\begin{itemize}
\item \textbf{Sparsity property:} $|\calI(\Z)|$ is small compared to $n$.
\item \textbf{Screening property:} $\phi_o \in
  \Phi_{\calI(\Z)},\;\textrm{if}\; \Z\; \textrm{generated according
    to}\;\phi_o$.
\end{itemize}
Formulating the sparsity property precisely, i.e. specifying the rate at which the size of the active set $\calI(\Z)$
  can grow with $n$, is a research topic in its own right (see also
  the discussion in Section~\ref{sec:discussion}). We do not address
  this topic here.
The screening property guarantees that the model $\{d(\cdot|x;\phi):\phi
\in \Phi_{\calI(\Z)}\}$, selected by $\calI$, is correctly specified in the sense that it contains the true density function $d(\cdot|x;\phi_o)$.
L1-penalized likelihood methods \citep{fan01variable} serve as our
prime example of screening procedures. Consider estimators of the form
\begin{eqnarray}\label{eq:penmle}
&\hat{\phi}_{\lambda}=\argmax\limits_{\phi\in\Phi}\ell(\phi;\Y|\X)-\lambda\|\phi\|_1,
\end{eqnarray}
where $\ell(\phi,\Y|\X)$ denotes the conditional log-likelihood and $\lambda$ is a
non-negative regularization parameter. In the context of linear regression
(\ref{eq:penmle}) coincides with the Lasso estimator
introduced by \cite{tibshirani96regression}. Other important example
of the form~(\ref{eq:penmle}) include the elastic net
\citep{zou05regularization}, the grouped Lasso \citep{yuan06model},
the graphical Lasso \citep{friedman2007sic} or the Lasso for
generalized linear models \citep{Park2007}. Estimators of the form~(\ref{eq:penmle}) have
been shown to be extremely powerful: they are suitable for
high-dimensional data and lead to sparse solutions. A
screening procedure can be defined by setting
\begin{eqnarray}\label{eq:penscreen}
  &\calI_{\lambda}(\Z)=\{j: \hat{\phi}_{\lambda,j}\neq 0\}.
\end{eqnarray}
The question is whether or not $\calI_{\lambda}(\Z)$ satisfies both
the sparsity \emph{and} the screening
property. In principle the answer depends on the choice of the
tuning parameter $\lambda$. Too strong regularization (too large $\lambda$) leads to
very sparse solutions, but with relevant parameter components incorrectly set
to zero, whereas too little regularization (too small
$\lambda$) results in too large active sets.
Common practice is to 
choose $\lambda$ in a prediction optimal manner, for example by
optimizing a cross-validation score. This gives typically sparse
solutions with a larger number of non-zero components than the true
number \citep{meinshausen04consistent}. It seems to be reasonable in practice to
assume that the sparsity and
screening property hold for the procedure $\calI_{\lambda}(\cdot)$ obtained via
$\ell_1$-regularization. In fact, \cite{buhlmann2012}
points out that the screening property is a very useful concept for
$\ell_1$-regularization which holds
under much milder assumptions than consistency in variable selection.

By applying the screening procedure $\calI_{\lambda}$ on $\U_{\rm in}$ and $\V_{\rm in}$
separately, we obtain active-sets
 $$I_u=\calI_{\lambda}(\U_{\rm in}), \quad I_v=\calI_{\lambda}(\V_{\rm in}).$$ 
Further we obtain active-set $$I_{uv}=\calI_{\lambda}\big((\U_{\rm in},\V_{\rm
  in})\big)$$ by applying $\calI$ jointly on $(\U_{\rm in},\V_{\rm
  in})$. For the rest of this section we treat the active-sets $I_u$, $I_v$ and $I_{uv}$ as fixed, i.e., not depending on the sample size.


\subsection{Restricted likelihood ratio statistic} \label{sec:restricted.loglik}
We define model $M_{\rm joint}$ with shared
parameter space $\Phi_{I_{uv}}$ for both population $U$ and $V$ jointly as

$$M_{\rm joint}=\Big\{d(y_u|x_u;\phi_{uv})d(y_v|x_v;\phi_{uv}):\phi_{uv} \in
\Phi_{I_{uv}}\Big\}$$
 and model $M_{\rm ind}$ with individual parameter spaces
 $\Phi_{I_u}$ and  $\Phi_{I_v}$ for each population as

$$M_{\rm ind}=\Big\{d(y_u|x_u;\phi_u)d(y_v|x_v;\phi_v):(\phi_u,\phi_v) \in
\Phi_{I_u}\times\Phi_{I_v}\Big\}.$$

The log-likelihood functions with respect to models $M_{\rm joint}$
and $M_{\rm ind}$ are given by
\begin{eqnarray*}
  \label{eq:loglik}
  L_{n_u,n_v}^{\rm joint}(\phi_{uv})&=&\sum_{i=1}^{n_u}\log
  d(\Y_{u,i}|\X_{u,i};\phi_{uv})+\sum_{i=1}^{n_v}\log d(\Y_{v,i}|\X_{v,i};\phi_{uv})\\
L_{n_u,n_v}^{\rm ind}(\phi_u,\phi_v)&=&\sum_{i=1}^{n_u}\log
  d(\Y_{u,i}|\X_{u,i};\phi_u)+\sum_{i=1}^{n_v}\log d(\Y_{v,i}|\X_{v,i};\phi_v).
\end{eqnarray*}


We will test hypothesis~(\ref{eq:h0}) based on the \emph{restricted}
log-likelihood ratio test-statistic defined as 
\begin{eqnarray}\label{eq:lrt}
\textrm{LR}_{n_u,n_v}&=&2\{\max L_{n_u,n_v}^{\rm ind}(\phi_u,\phi_v)-\max L_{n_u,n_v}^{\rm
    joint}(\phi)\}\\
&=&2\{L_{n_u,n_v}^{\rm ind}(\hat\phi_u,\hat\phi_v)-L_{n_u,n_v}^{\rm
    joint}(\hat\phi_{uv})\},\nonumber
\end{eqnarray}
where $(\hat\phi_u,\hat\phi_v)$ and $\hat\phi_{uv}$ denote the
maximum likelihood estimators corresponding to models $M_{\rm
  ind}$ and $M_{\rm
  joint}$. 

It is crucial to note that testing based on (\ref{eq:lrt}) is non-trivial and
involves non-nested model comparison \citep{vuong1989}. The difficulty arises from the
fact that model $M_{\rm joint}$ is not nested in $M_{\rm ind}$, or in other
terms $\Phi_J {\not \subset}
\;\Phi_{I_u}\cup\Phi_{I_v}$. Non-nestedness of these models is
fundamental and is not only an artefact of the random nature of
the screening
procedure. If $\phi_u \neq \phi_v$, than $\calI$ applied on two
different populations mixed together will set different components of
$\phi$ to zero than $\calI$ applied two both populations individually.
This is a consequence of model misspecification and is also well known
as Simpson's paradox in which an association present in two different
groups can be lost or even reversed when the groups are combined.

Under $\mathbf{H_A}$, or if $\phi_u\neq\phi_v$, then we have:
\begin{prop}[]\label{prop:lrtHA}
Assume that the screening property holds and consider the sets $I_u,
I_v$ and $I_{uv}$ as fixed. Then under {$\mathbf{H_A}$} and regularity
assumptions (A1)-(A3) (listed in Appendix~\ref{sec:app.proof}):
$$\textrm{LR}_{n_u,n_v}=2\{L_{n_u,n_v}^{\rm ind}(\hat\phi_u,\hat\phi_v)-L_{n_u,n_v}^{\rm
    joint}(\hat\phi_{uv})\}\overset{a.s.}{\rightarrow}  \infty\quad (n_u,n_v \rightarrow \infty).$$
\end{prop}
A proof is given in Appendix~\ref{sec:app.proof}. Based on Proposition~\ref{prop:lrtHA} we reject {$\mathbf{H_0}$} if
$\textrm{LR}_{n_u,n_v}$ exceeds some critical value. The critical value is
chosen to control the type-I error at some level of
significance $\alpha$. Alternatively, one may directly compute a p-value. 
In order to compute a critical value and/or determine a
p-value we obtain in
Section~\ref{sec:asymptotic} the asymptotic distribution of
$\textrm{LR}_{n_u,n_v}$ under $\mathbf{H_0}$. 





\subsection{Asymptotic Null Distribution}\label{sec:asymptotic}
The work of \cite{vuong1989} specifies the asymptotic distribution
of the log-likelihood ratio statistic for comparing two competing
models in a very general setting, in particular Vuong's theory covers the case where the two models are non-nested. We apply Theorem~3.3 of \cite{vuong1989} to the special case where the competing models are $M_{\rm joint}$ and $M_{\rm ind}$.


Let $\phi^*_{uv}$ and $(\phi^*_u,\phi^*_v)$ be pseudo-true
values of model $M_{\rm joint}$, respectively $M_{\rm ind}$:
\begin{eqnarray*}
  &\phi^*_{uv}=\argmin\limits_{\phi\in\Phi_{I_{uv}}}\left(\E[\E_{\phi_u}[\log
  d(Y_u|X_u;\phi)]]+\E[\E_{\phi_v}[\log d(Y_v|X_v;\phi)]]\right),\\
&\phi^*_{u}=\argmin\limits_{\psi\in\Phi_{I_{u}}}\E[\E_{\phi_u}[\log
  d(Y_u|X_u;\psi)]],\;\phi^*_{v}=\argmin\limits_{\xi\in\Phi_{I_{v}}}\E[\E_{\phi_v}[\log
  d(Y_v|X_v;\xi)]].
\end{eqnarray*}
Define for $a, b\in\{u,v,uv\}$, $c\in\{u,v\}$ and sets $A, B \subset \{1,\ldots,p\}$
\begin{eqnarray*}
  s_{A}(y|x;\phi)&=&\frac{\partial }{\partial \phi_{A}}\log
d(y|x;\phi)\quad\textrm{and}\\
B^c_{A B}(\phi^*_a;\phi^*_b)&=&\E[\E_{\phi_c}[s_{A}(Y|X;\phi^*_a)s_{B}(Y|X;\phi^*_b)^T]].
\end{eqnarray*}

We further consider the matrices
\begin{eqnarray*}
&B_{M_{\rm ind}}(\phi^*_u,\phi^*_v)=\begin{pmatrix}
  B^u_{I_u}(\phi^*_u)& 0\\
  0&  B^v_{I_v}(\phi^*_v)\\
\end{pmatrix},\quad B_{M_{\rm joint}}(\phi^*_{uv})=B^u_{I_{uv}}(\phi^*_{uv})+B^v_{I_{uv}}(\phi^*_{uv})
\end{eqnarray*}
and $$B_{M_{\rm joint} M_{\rm
    ind}}(\phi^*_{uv};\phi^*_u,\phi^*_v)=\left(B^u_{I_{uv}I_u}(\phi^*_{uv};\phi^*_u),B^v_{I_{uv}I_v}(\phi^*_{uv};\phi^*_v)\right).$$
The following theorem establishes the asymptotic distribution of $\textrm{LR}_{n_u,n_v}$.
\begin{theorem}[Asymptotic null-distribution of restricted likelihood
  ratio-statistic.]\label{thm:asy.null}
Assume that the screening property holds and consider the sets $I_u$, $I_v$
and $I_{uv}$ as fixed. If $\phi_u=\phi_v(=:\bar\phi)$ and under
regularity assumptions (A1)-(A6) (listed in Appendix~\ref{sec:app.proof}) we
have:
$$\textrm{LR}_{n_u,n_v}\overset{d}{\rightarrow} \Psi_r(\cdot;\nu),$$ with
$r=|I_u|+|I_v|+|I_{uv}|$ and $\Psi_r(\cdot;\nu)$  denotes the
distribution function of a weighted sum of $r$ independent chi-square
distributions where the weights
$\nu_j,\;j=1,\ldots,r$ are
eigenvalues of the matrix 
\begin{eqnarray}
\label{eq:wmatrix}
W&=&\begin{pmatrix}
  \ID_{r_u+r_v}& B_{M_{\rm ind} M_{\rm joint}} B^{-1}_{M_{\rm joint}}\\
   B_{M_{\rm joint} M_{\rm ind}}B^{-1}_{M_{\rm ind}}&  -\ID_{r_{uv}}\\

 \end{pmatrix}
\end{eqnarray}
\end{theorem}

A proof is given in Appendix~\ref{sec:app.proof}. Set
$J=I_{uv}\cap I_u\cap I_v$, $\mathring{I}_a=I_a-J$,
$\mathring{I}_b=I_b-J$ and $$Q^c_{\mathring{I}_a
  \mathring{I}_b}(\phi^*_a,\phi^*_b)=B^c_{\mathring{I}_a \mathring{I}_b}(\phi^*_a,\phi^*_b)-B^c_{\mathring{I}_a J}(\phi^*_a,\phi^*_b)B^c_{JJ}(\phi^*_a,\phi^*_b)^{-1}B^c_{J \mathring{I}_b}(\phi^*_a,\phi^*_b).$$ 
If we assume that the screening property holds then model $M_{\rm
  ind}$ is correctly specified and the pseudo-true values
$(\phi^*_u,\phi^*_v)$ equal the true values
$(\phi_u,\phi_v)$. Furthermore, if we have $\phi_u=\phi_v(=:\bar{\phi})$ then the screening property
guarantees that also $M_{\rm joint}$ is correctly specified and that
$\phi^*_{uv}=\bar{\phi}$. We then write 

$$B^c_{A
  B}(\phi^*_a;\phi^*_b)=B_{A
  B}\quad\textrm{and}\quad Q^c_{\mathring{I}_a
  \mathring{I}_b}(\phi^*_a,\phi^*_b)=Q_{\mathring{I}_a
  \mathring{I}_b}.$$ 
In Appendix~\ref{sec:app.proof} we prove the following
proposition which characterizes the weights $\nu_j$ defined as eigenvalues of the matrix $W$ in Theorem~\ref{thm:asy.null}.


\begin{prop}[Characterization of eigenvalues]\label{prop:evals}
The eigenvalues $\nu_j$, $j=1,\ldots|I_u|+|I_v|+|I_{uv}|$, of matrix W in Theorem~\ref{thm:asy.null} can be characterized as follows:

If $|I_u|+|I_v|>|I_{uv}|$:
\begin{itemize}
\item $2\;|J|$ eigenvalues are 0.
\item $|I_u|+|I_v|-|I_{uv}|$ eigenvalues are 1.
\item The remaining eigenvalues equal $\pm\sqrt{1-\mu_j}$, where $\mu_j$ are eigenvalues of 
  \begin{eqnarray}
    \label{eq:propeq1}
    &\big(Q_{\minI_{uv}\minI_u}Q^{-1}_{\minI_u}Q_{\minI_u\minI_{uv}}+Q_{\minI_{uv}\minI_v}Q^{-1}_{\minI_v}Q_{\minI_v\minI_{uv}}\big)(2Q_{\minI_{uv}})^{-1}.
  \end{eqnarray}
\end{itemize}
If $|I_{uv}|>|I_u|+|I_v|$
\begin{itemize}
\item $2\;|J|$ eigenvalues are 0.
\item $|I_{uv}|-(|I_u|+|I_v|)+|J|$ eigenvalues are -1.
\item $|J|$ eigenvalues are +1.
\item The remaining eigenvalues equal $\pm\sqrt{1-\mu_j}$, where $\mu_j$ are eigenvalues of
  \begin{eqnarray}
    \label{eq:propeq2}
    &\begin{bmatrix}(Q_{\minI_u\minI_{uv}}(2Q_{\minI_{uv}})^{-1}Q_{\minI_{uv}\minI_u})Q^{-1}_{\minI_u}&(Q_{\minI_u\minI_{uv}}(2Q_{\minI_{uv}})^{-1}Q_{\minI_{uv}\minI_v})Q^{-1}_{\minI_v}\\
(Q_{\minI_v\minI_{uv}}(2Q_{\minI_{uv}})^{-1}Q_{\minI_{uv}\minI_u})Q^{-1}_{\minI_u}&(Q_{\minI_v\minI_{uv}}(2Q_{\minI_{uv}})^{-1}Q_{\minI_{uv}\minI_v})Q^{-1}_{\minI_v}
\end{bmatrix}.
  \end{eqnarray}

\end{itemize}
\end{prop}
Expressions~(\ref{eq:propeq1}) and (\ref{eq:propeq2}) of
 Proposition~\ref{prop:evals} have nice interpretations in terms of analyzing the variances of the models
  $M_{\rm ind}$ and $M_{\rm joint}$. Consider the random
variables $$Z^c=\frac{1}{\sqrt{n}}\sum\limits_{i=1}^{n}s(\Y_{c,i}|\X_{c,i};\bar\phi)\;(c\in \{u,v\}),$$
which are asymptotically Normal distributed with mean zero. 
If $\mathbf{res}^c_{I_a}$ and $\mathbf{res}^c_{I_b}$ denote the residuals obtained from
regressing $Z^c_{I_a}$ and $Z^c_{I_b}$
against $Z^c_J$, then $Q_{\mathring{I}_a
  \mathring{I}_b}$ is the covariance between these residuals.
It is easy to see
that the matrix
$Q_{\minI_{uv}\minI_u}Q^{-1}_{\minI_u}Q_{\minI_u\minI_{uv}}$
equals $$\Var(\mathbf{res}^c_{I_{uv}})-\Var(\mathbf{res}^c_{I_{uv}}|\mathbf{res}^c_{I_{u}}),$$
and therefore expression~(\ref{eq:propeq1}) of Proposition~\ref{prop:evals} can be interpreted as the asymptotic
variance of model $M_{\rm joint}$
not explained by model $M_{\rm ind}$. Similarly, we can see that (\ref{eq:propeq2}) describes the variance of model $M_{\rm ind}$ not
explained by $M_{\rm joint}$. 

We further point out two special cases of Proposition~\ref{prop:evals}: If $M_{\rm joint}$ is nested in $M_{\rm ind}$, i.e., $|\minI_{uv}|=0$,
  then $\textrm{LR}_{n_u,n_v}$ follows asymptotically a chi-squared
  distribution with degrees of freedom equal $|I_u|+|I_v|-|I_{uv}|$.
If $|\minI_u|=|\minI_v|=0$ then $\textrm{LR}_{n_u,n_v}$ is
  asymptotically distributed according to $\chi^2_{|J|}-\chi^2_{|I_{uv}|-|J|}$.

\subsection{Estimation of the weights $\nu$ in $\Psi_r(\cdot;\nu)$}\label{sec:weights}
In practice the weights $\nu$ of the weighted sum of chi-squared null distribution have to be
estimated from the data. In light of Theorem~\ref{thm:asy.null}, it would be straightforward to estimate the quantities $B^c_{I_a I_b}(\phi^*_a,\phi^*_b)$ and plug them into expression (\ref{eq:wmatrix}) to obtain an estimate $\hat W$. Estimating $\nu$ then involves computation of
$r=|I_u|+|I_v|+|I_{uv}|$ eigenvalues. Despite model reduction in the screening step, $r$ can be a rather large number and can result in inefficient and inaccurate estimation. However, if both populations arise from the same distribution, then the overlap
of the active-sets $J=I_{uv}\cap
I_u\cap I_v$ is large compared to $I_u$, $I_v$ and
$I_{uv}$. According to Proposition~\ref{prop:evals}, the number of
eigenvalues $s=\min\{|I_{uv}|-|J|,|I_u|+|I_v|-|J|\}$ which remain to be
estimated is small compared to $r$. We therefore estimate matrix~(\ref{eq:propeq1}) by
\begin{eqnarray}
  \label{eq:estQ1}
  \big(\hat Q^u_{\minI_{uv}\minI_u}(\hat Q^{u}_{\minI_u})^{-1}\hat Q^u_{\minI_u\minI_{uv}}+\hat Q^v_{\minI_{uv}\minI_v}(\hat Q^v_{\minI_v})^{-1}\hat Q^v_{\minI_v\minI_{uv}}\big)(\hat Q^u_{\minI_{uv}}+\hat Q^v_{\minI_{uv}})^{-1},
\end{eqnarray}

and similarly matrix~(\ref{eq:propeq2}) by 
\begin{eqnarray}
  \label{eq:estQ2}
  \begin{bmatrix}(\hat Q^u_{\minI_u\minI_{uv}}(\hat
  Q^u_{\minI_{uv}}+\hat Q^v_{\minI_{uv}})^{-1}\hat
  Q^u_{\minI_{uv}\minI_u})(\hat Q^u_{\minI_u}) ^{-1}&(\hat Q^u_{\minI_u\minI_{uv}}(\hat Q^u_{\minI_{uv}}+\hat Q^v_{\minI_{uv}})^{-1}\hat Q^v_{\minI_{uv}\minI_v})(\hat Q^v_{\minI_v})^{-1}\\
(\hat Q^v_{\minI_v\minI_{uv}}(\hat Q^u_{\minI_{uv}}+\hat Q^v_{\minI_{uv}})^{-1}\hat Q^u_{\minI_{uv}\minI_u})(\hat Q^u_{\minI_u})^{-1}&(\hat Q^v_{\minI_v\minI_{uv}}(\hat Q^u_{\minI_{uv}}+\hat Q^v_{\minI_{uv}})^{-1}\hat Q^v_{\minI_{uv}\minI_v})(\hat Q^v_{\minI_v})^{-1}
\end{bmatrix}.
\end{eqnarray}

Here, $\hat Q^c_{\minI_{a}\minI_b}=\hat{B}^c_{\minI_a \minI_b}-\hat{B}^c_{\minI_a J}\hat{B}^c_{J J}\hat{B}^c_{J \minI_b}$ and $\hat{B}^c_{I_a I_b}$ denotes a consistent estimator of $B^c_{I_a I_b}(\phi^*_a,\phi^*_b)$. One possibility is to use the sample analogues:
\begin{eqnarray*}
  \hat{B}^c_{I_a I_b,\textrm{sample}}&=&\frac{1}{n_c}\sum\limits_{i=1}^{n_c}s_{I_a}(\Y_{c,i}|\X_{c,i};\hat{\phi}_a)\;s_{I_b}(\Y_{c,i}|\X_{c,i};\hat{\phi}_b)^T.
\end{eqnarray*}
Another way is to plug-in estimators $\hat\phi_u$ and
$\hat\phi_v$ into the expectation with respect to $\phi_u$ and
$\phi_v$:
\begin{eqnarray}\label{eq:plugin}
  \hat{B}^c_{I_a I_b,\textrm{plug-in}}&=&\frac{1}{n_c}\sum\limits_{i=1}^{n_c}\E_{\hat\phi_c}[s_{I_a}(Y|\X_{c,i};\hat\phi_a)\;s_{I_b}(Y|\X_{c,i};\hat\phi_b)^T].
\end{eqnarray}

\begin{figure}[htbp!]
\begin{centering}
\includegraphics[scale=0.48]{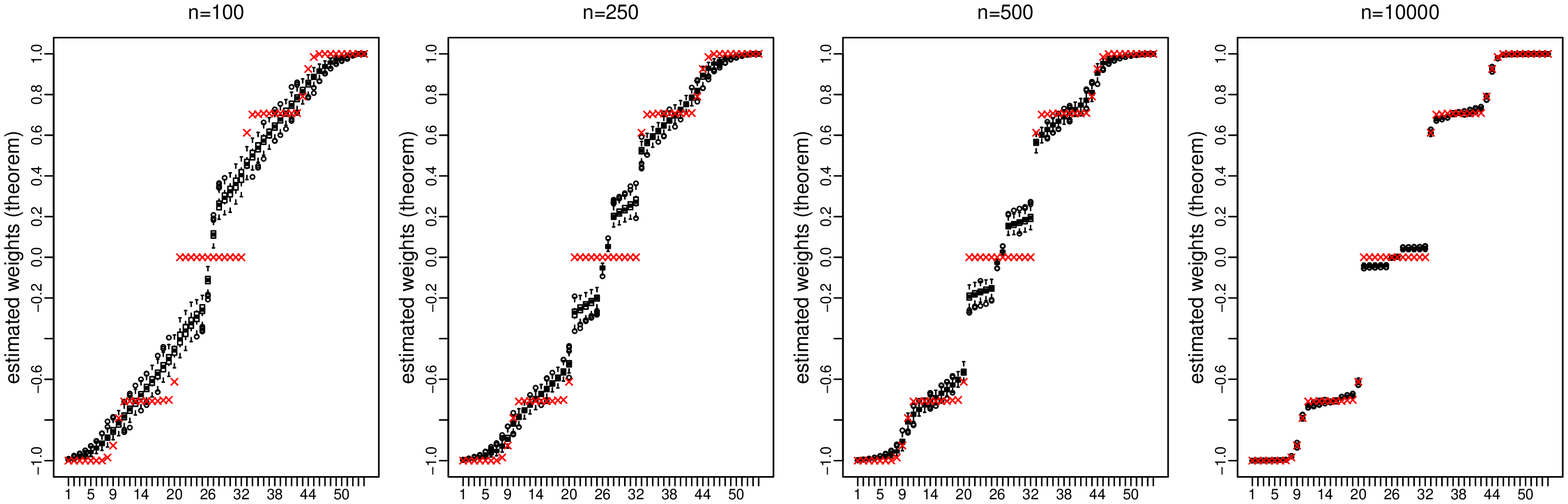} 
\includegraphics[scale=0.48]{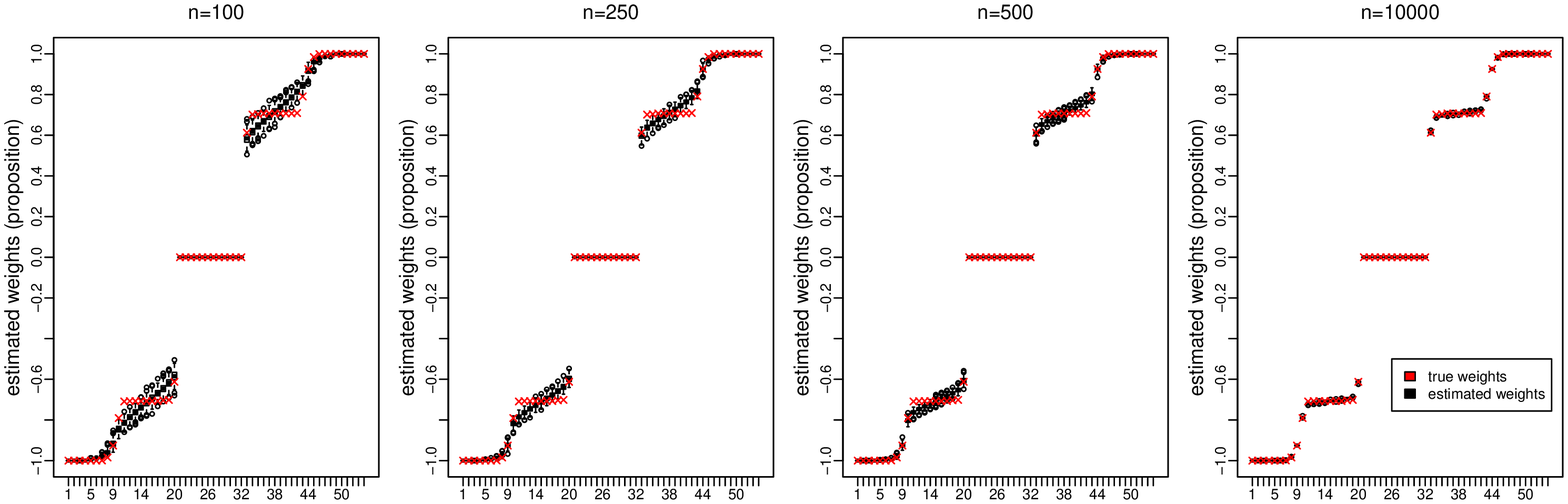} 
  \caption[]
  {Comparison of estimated weights required for null
    distribution (boxplots) with true values (in red) for a regression example. Upper row shows estimates obtained by directly
    computing the eigenvalues of $\hat{W}$, Theorem~\ref{thm:asy.null}. Lower row shows estimates
    obtained via Proposition~\ref{prop:evals} and equations (\ref{eq:estQ1})-(\ref{eq:estQ2}). [Linear
regression model with regression coefficients $\beta\in \R^{100}$,
$\beta_j=1 \;(j=1,\ldots,5)$ and zero elsewhere, $\sigma^2=1$. $I_{u}, I_{v}$
and $I_{uv}$ obtained using sampled data of size $(n:=)n_u=n_v=100$. Eigenvalues computed using sampled data of size $(n:=)n_u=n_v=100, 250, 500,
10000$.]}
\label{fig:eval}
\end{centering}
\end{figure}

\begin{figure}[htbp!]
\begin{centering}
   \includegraphics[scale=0.55]{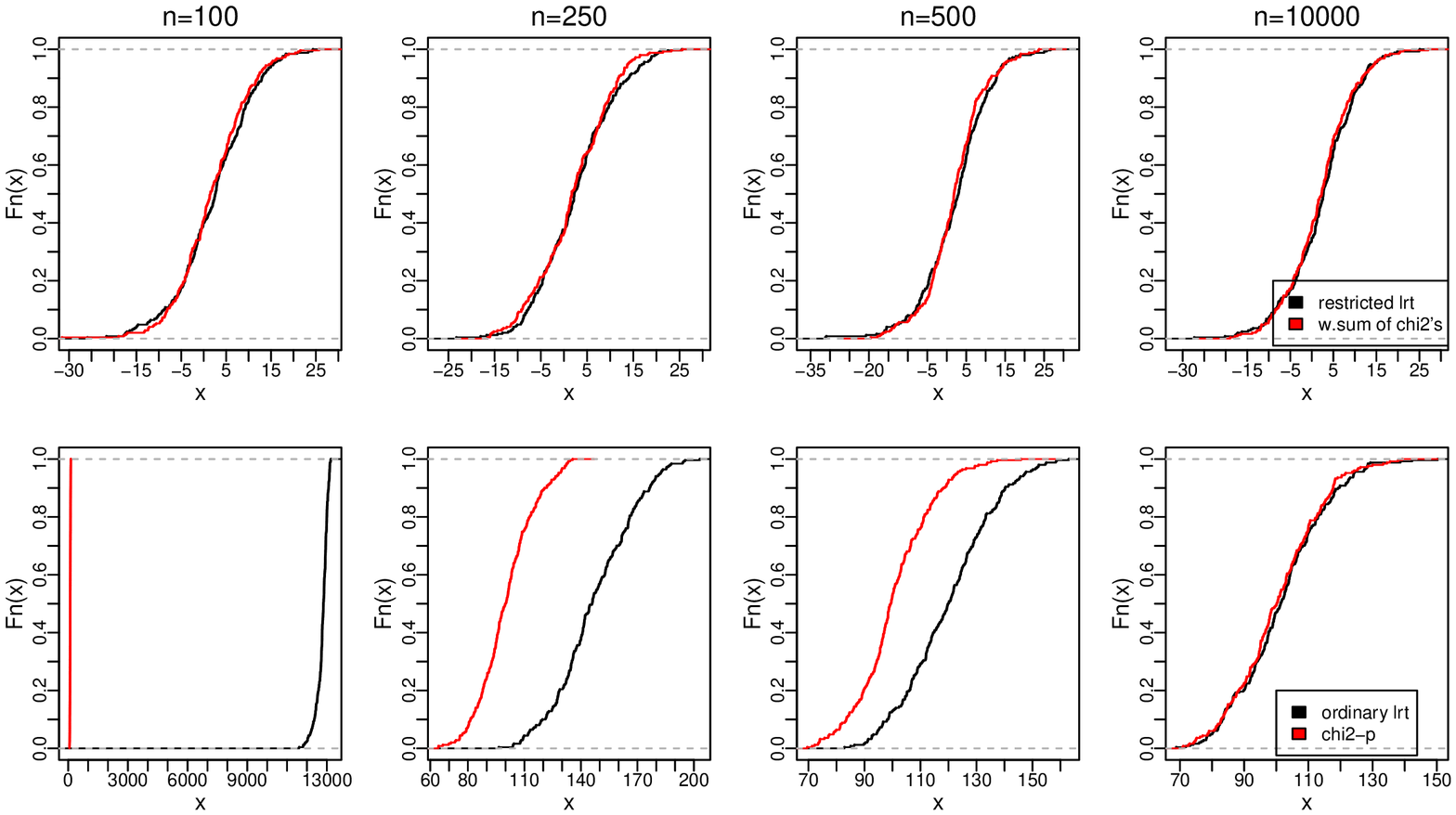} 
  \caption[]
  {Distribution functions for ordinary and restricted likelihood-ratio
    statistic and asymptotic approximations. Upper row of panels: approximation of restricted
    likelihood-ratio (black) by the weighted sum of chi-squared distribution (red). Lower
    row of panels: approximation of ordinary likelihood-ratio (black) by
    $\chi^2_p$ distribution (red).  [A linear
regression model with $l=100$ predictors (for more details see caption
of Figure~\ref{fig:eval}).]} 
\label{fig:distr_lr}
\end{centering}
\end{figure}

Figure~\ref{fig:eval} shows for a
linear regression example with $l=100$ predictors 
the estimated weights $\hat{\nu}$ (upper panels show $\hat{\nu}$ obtained
using directly Theorem~\ref{thm:asy.null}; lower panels show
$\hat{\nu}$ obtained via Proposition~\ref{prop:evals}). 
Estimating the eigenvalues with help of Proposition~\ref{prop:evals}
works better than direct computation according to
Theorem~\ref{thm:asy.null}. In particular the
true zero eigenvalues are poorly estimated with the direct approach. Figure~\ref{fig:distr_lr}
illustrates for the same example the quality of approximation of the
ordinary and restricted likelihood-ratio with their asymptotic
counterparts. The weighted sum of chi-squares approximates well already for small sample sizes, whereas
the $\chi^2_p$ comes close to the distribution function
of the ordinary likelihood-ratio only for very large $n$'s.

\subsection{P-Values and Multi-Splitting}\label{sec:multisplit.pval}
In this section we demonstrate how to obtain p-values for testing
hypothesis~(\ref{eq:h0}) using the methodology developed in
Sections~\ref{sec:screen}-\ref{sec:weights}. The basic workflow is to split the data into
two parts, do screening on one part and derive asymptotic p-values on the other
part. A p-value computed based on a one-time single split depends heavily on the arbitrary choice of the split and amounts to a ``p-value lottery'' \citep{meinshausenpval2009}: for finite samples and for some specific split, the screening or the sparsity property might be violated which then results in a erroneous p-value. An alternative to a single arbitrary sample split is to split the data repeatedly. \cite{meinshausenpval2009} demonstrated in the context of variable selection in the high-dimensional regression model that such a multi-split approach gives improved and more reproducible results. We adopt this idea and divide the data repeatedly into $k=1,\ldots,K$ different splits. Let
($\U^k_{\rm in},\V^k_{\rm in}$) and ($\U^k_{\rm out},\V^k_{\rm out}$) be
the first and second half of split $k$. On the first half screening is
performed by solving three times the $\ell_1$-regularized log-likelihood problem
(\ref{eq:penmle}), individually for each population and for both populations
pooled together. The tuning parameter $\lambda$ is always chosen by
cross-validation. This gives models $M^k_{\rm joint}$ and $M^k _{\rm
  ind}$ defined via active-sets $\calI_{\lambda_{\rm cv}}(\U^k_{\rm
  in}),\; \calI_{\lambda_{\rm cv}}(\V^k_{\rm in})$ and $\calI_{\lambda_{\rm cv}}\left((\U^k_{\rm in},\V^k_{\rm in})\right)$. Then, the
restricted log-likelihood ratio $\textrm{LR}^k$ is evaluated on the
second half of split $k$ and a one-sided p-value is computed according
to
\begin{eqnarray*}
 \mathcal{P}^k&=&1-\Psi_r(\textrm{LR}^k;\hat \nu),
\end{eqnarray*}
where $r=|\calI_{\lambda_{\rm cv}}(\U^k_{\rm
  in})|+|\calI_{\lambda_{\rm cv}}(\V^k_{\rm
  in})|+|\calI_{\lambda_{\rm cv}}\left((\U^k_{\rm in},\V^k_{\rm in})\right)|$ and $\Psi_r(\cdot;\cdot)$ is defined in Theorem~\ref{thm:asy.null}. The weights $\hat\nu$ are estimated
from the second half of split $k$ as described in Section~\ref{sec:weights}. As in \cite{meinshausenpval2009} we aggregate p-values
$\mathcal{P}^k, k=1,\ldots,K$, obtained from all different splits using
the formula:

$$\mathcal{P}_{\rm agg}=\min\left((1-\gamma_{\rm min})\!
\inf\limits_{\gamma \in (\gamma_{\rm min},1)}
q_{\gamma}\left(\{\mathcal{P}^b/\gamma;b=1,\ldots,B\}\right),1\right),$$

where $q_{\gamma}(\cdot)$ is the empirical $\gamma$-quantile
function. This procedure is summarized in Algorithm~\ref{alg:multisplit}. We refer to the choice $K=1$ as the single-split method and to the choice $K>1$ as the multi-split method.

\begin{algorithm}                      
\textbf{Input} number of splits $K$.
\begin{algorithmic}[1]         
\For{$k=1,\ldots,K$} \medskip 
\State Randomly split data into ($\U^{k}_{\rm in}$, $\U^k_{\rm out}$) and
($\V^k_{\rm in}$, $\V^k_{\rm out}$). 
\medskip \medskip
\State \emph{Screening} on ($\U^k_{\rm in},\V^k_{\rm in}$): \medskip

Compute active-sets $I^k_u\leftarrow\calI_{\lambda_{\rm cv}}(\U^{k}_{\rm in})$;
$I^k_v\leftarrow\calI_{\lambda_{\rm cv}}(\V^{k}_{\rm in})$;
$I^k_{uv}\leftarrow\calI_{\lambda_{\rm cv}}(\U^{k}_{\rm
  in},\V^{k}_{\rm in})$.

\medskip
[$\calI_{\lambda}(\cdot)$: defined in (\ref{eq:penscreen}),
$\lambda_{\rm cv}$ obtained by cross-validation]
\medskip

\medskip
\State \emph{Significance Testing} on ($\U^k_{\rm out},\V^k_{\rm
  out}$): 

\medskip
Compute test-statistic:
$\textrm{LR}^k\leftarrow 2\{L_{n_u,n_v}^{\textrm{ind}^k}(\hat\phi_u,\hat\phi_v)-L_{n_u,n_v}^{\textrm{joint}^k}(\hat\phi_{uv})\}$

\medskip
Compute p-value: $\mathcal{P}^k\leftarrow
1-\Psi_r(\textrm{LR}^k;\hat\nu)$

\medskip[$L_{n_u,n_v}^{\textrm{ind}^k}\!(\hat\phi_u,\!\hat\phi_v),
L_{n_u,n_v}^{\textrm{joint}^k}\!(\hat\phi_{uv}\!)$: defined in
(\ref{eq:lrt}), $\hat\nu$ obtained as described in
Section~\ref{sec:weights}]\medskip
\EndFor
\medskip
\end{algorithmic}
\textbf{Output} 

\hspace{0.5cm}\textbf{if} {$K=1$}: {$\mathcal{P}\leftarrow\mathcal{P}^1$}.

\hspace{0.5cm}\textbf{if} {$K>1$}: {$\mathcal{P}_{\rm agg}\leftarrow
\min\left((1-\gamma_{\rm min})\!
\inf\limits_{\gamma \in (\gamma_{\rm min},1)} q_{\gamma}\left(\{\mathcal{P}^k/\gamma;k=1,\ldots,K\}\right),1\right)$}.

\caption{Single- and multi-split algorithm for high-dimensional two-sample testing}  \label{alg:multisplit} 
\end{algorithm}

\section{Examples}\label{sec:regr.ggm}
As mentioned in the beginning of Section~\ref{sec:method} two examples
of our approach are \emph{differential regression} where the
populations follow linear regression models and \emph{differential
  network} where the populations are generated from Gaussian graphical
models. In this Section we provide details on both examples.

\paragraph{Differential regression} Consider a regression model
\begin{eqnarray}
  \label{eq:regr}
  Y=X\beta+\epsilon,
\end{eqnarray}
with $Y\subset\R$ ($k=1$), $X\subset \R^{l}$ ($l=p-1$) and random error $\epsilon\sim\calN(0,\sigma^2)$. 
With $\phi=(\beta,\sigma^2)\in \R^p$ 
the score function is given by
\begin{eqnarray*}
s(y|x;\phi)&=&\left(\frac{(y-\beta^T x)
    x}{\sigma^2};\frac{1}{2\sigma^{2}}\left(\frac{(y-\beta^T x)^2}{\sigma^2}-1\right)\right).
\end{eqnarray*}
In Appendix~\ref{sec:app.expect} we show that
\begin{eqnarray}\label{eq:betamat.regr}
 \E_{\phi_c}[s_A(Y|X;\phi_a)\;s_B(Y|X;\phi_b)^T]&=&\frac{{X_A}X_B^T}{{\sigma_a}^2
    {\sigma_b}^2}\left(\sigma^2_c+(\beta_c-\beta_a)^T X X^T(\beta_c-\beta_b)\right).
\end{eqnarray}
Now, given data $\U=(\Y_u,\X_u)$ and $\V=(\Y_v,\X_v)$ we obtain
p-values by following  Algorithm~1 (see
Section~\ref{sec:multisplit.pval}). For the regression model,
$\ell_1$-penalized maximum likelihood estimation, used in the screening
step, coincides with the Lasso \citep{tibshirani96regression} and is
implemented in the \textbf{R}-package \texttt{glmnet} \citep{friedman2010}. With the help of formula~(\ref{eq:betamat.regr}) we
can easily compute plug-in estimates $\hat{B}^c_{I_a
  I_b,\textrm{plug-in}}$ (see equation~(\ref{eq:plugin})) and then the
weights $\hat\nu$ of the asymptotic null distribution are obtained as outlined in
Section~\ref{sec:weights}. We use the algorithm of \cite{davies1980},
implemented in the \textbf{R}-package \texttt{CompQuadForm} \citep{duchesne2012}, to compute the distribution function of $\Psi_r(\cdot;\hat\nu)$.






\paragraph{Differential network}
Let $Y\subset \R^k$ ($l=0$) be Gaussian distributed with zero mean and covariance
$\Sigma$, i.e., $\calN(0,\Sigma)$. A Gaussian graphical model with
undirected graph $G$ is then defined by locations of zero entries in
the inverse covariance matrix $\Omega=\Sigma^{-1}$, i.e., $(j,j'){\not \in} G
\Leftrightarrow \Omega_{jj'}=0$. Setting
$\phi=(\{-0.5\Omega_{jj}\}_{j=1}^k,\{-\Omega_{jj'}\}_{j> j'})\in \R^p$ ($p=(k+1)k/2$), the score function is given by
$$s_{(j,j')}(Y;\phi)=Y^{(j)}Y^{(j')}-\Sigma_{jj'}.$$ By invoking
formulas on fourth moments of a multivariate normal
distribution (see Appendix~\ref{sec:app.expect}) we find:
\begin{eqnarray}\label{eq:betamat.ggm}
 \E_{\phi_c}[s_{(j,j')}(Y;\phi_a)s_{(l,l')}(Y;\phi_b)]&=&\Sigma_{c,jj'}\Sigma_{c,ll'}+\Sigma_{c,jl}\Sigma_{c,j'l'}+\Sigma_{c,jl'}\Sigma_{c,j'l}\\\nonumber
&&-\Sigma_{a,jj'}\Sigma_{c,ll'}-\Sigma_{c,jj'}\Sigma_{b,ll'}+\Sigma_{a,jj'}\Sigma_{b,ll'}.
\end{eqnarray}
In the Gaussian graphical model case $\ell_1$-regularized maximum likelihood estimation is
well-known under the name \emph{Graphical Lasso} (or GLasso) and is implemented in the
\textbf{R}-package \texttt{glasso} \citep{friedman2007sic}. As in differential regression, given data $\U=\Y_u$ and $\V=\Y_v$,
plug-in estimates $\hat{B}^c_{I_a
  I_b,\textrm{plug-in}}$ are computed by
formula~(\ref{eq:betamat.ggm}), weights $\hat{\nu}$ are obtained
subsequently as described in Section~\ref{sec:weights} and for p-value
calculation we follow Algorithm 1.





\section{Numerical Results}\label{sec:exp.num}
\subsection{Simulations}\label{sec:exp.sim}

We consider the following simulation settings:
\begin{description}
\item[Setting 1]  Differential regression: Generate population $U=(Y_u,X_u)$ and $V=(Y_v,X_v)$
  according to regression model  (\ref{eq:regr}). $X_u$ and $X_v$ are
  generated according to $\calN(0,\Sigma)$ with
  $\Sigma_{jj'}=0.5^{|j-j'|}$. We set $(n:=)n_u=n_v=200$ and $l=7, 10, 25, 50, 100, 195$ ($l=p-1$) and choose $\sigma_u$ and $\sigma_v$ such that the signal-to-noise ratio (SNR) equals~10.

Under $\mathbf{H_0}$, the regression coefficients
$\bar\beta=\beta_u=\beta_v$ have
  5 non-zero elements (at random location) with values~1.

Under $\mathbf{H_A}$, the regression coefficients $\beta_u$
  and $\beta_v$ have entries of values~1 at three common locations and
  have entries of value $\alpha_1$ at two different locations. $\alpha_1=0.25, 0.5$.
\item[Setting 2]  As setting 1 but with $\mathrm{SNR}=5$.

\item[Setting 3] As setting 1 but as a predictor matrix $\X$ we use for
  both populations gene expression data from $n=594$ ovarian
  carcinomas. The data is publicly available at The Cancer Genome
  Atlas (TCGA) data portal (\url{http://www.cancergenome.nih.gov}). We
  select the $l=7, 10, 25, 50, 100, 250$ genes
  exhibiting the highest empirical variance among samples.  We choose
  $\sigma_u$ and $\sigma_v$ such that \textrm{SNR}=10.
 
\item[Setting 4] Differential network: Generate both population
  according to $Y~\sim\calN(0,\Sigma)$ with
  $(n:=)n_u=n_v=300$ and $k=5, 10, 25, 50, 75$. Note, that $Y\subset \R^k$ and $p=(k+1)k/2$.

Under $\mathbf{H_0}$, the inverse covariance matrices $\Omega_u$ and
$\Omega_v$ are equal and have $k$ non-zero entries at random
locations.

Under $\mathbf{H_A}$, the inverse covariance matrices $\Omega_u$ and
$\Omega_v$ have $\lceil k\alpha_2\rceil$ non-zero entries in common and
$k-\lceil k\alpha_2\rceil $ non-zero's at different locations,
$\alpha_2=0.5, 0.8$.

\end{description}
The values $\alpha_1$ and $\alpha_2$ control the strength of the
alternative for the regression and network cases respectively: A larger value $\alpha_1$ results in two regression
models which differ more. On the other hand a larger $\alpha_2$
signifies that the two networks have more edges in common and are therefore more
similar. For each of the four settings we perform $500$ simulation runs for the
null- and the different alternative hypothesis. We then compute for
each run a p-value with the \textbf{single-} and the \textbf{multi-split}
method (50 random splits, $\gamma_{\rm min}=0.05$). We compare
the proposed methodology with p-values obtained using
the asymptotic $\chi^2_{p}$-distribution of the ordinary
likelihood-ratio statistic. We call this latter approach \textbf{ordinary
  likelihood-ratio test}. Further we  compare with a {\bf permutation test} where we use
the symmetric Kullback-Leibler distance between 
$\ell_1$-regularized estimates as a test statistic. This
permutation test uses 100 random permutations and further
details are given in
Appendix~\ref{sec:app.perm}. For each setting we evaluate the fraction of
wrongly rejected null hypothesis (false positive rate) and the
fraction of correctly rejected hypothesis (true positive rate) at a
significance level of 5\%.

\begin{figure}[htbp!]
\begin{centering}
   \includegraphics[scale=0.65]{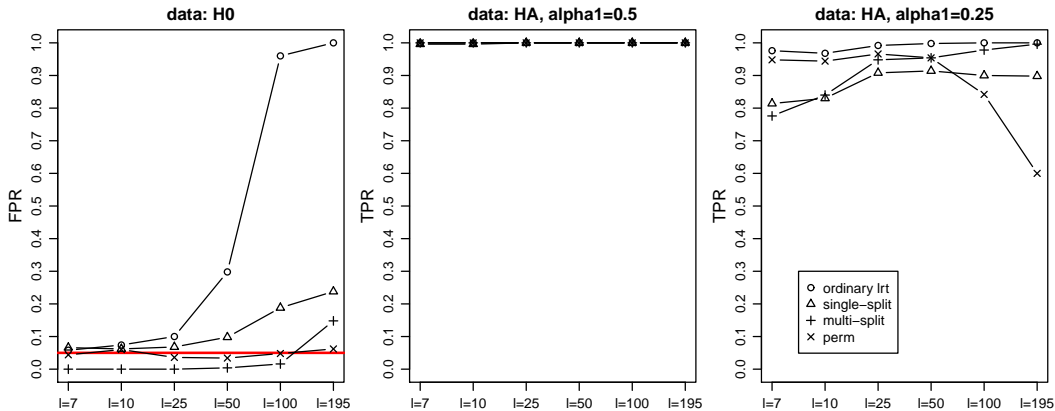} 
  \caption[]
  {Simulation study, Setting 1 (differential regression, SNR=10): False positive rate (FPR) and true positive rate
    (TPR) at 5\%-level shown as a function of number of predictors
    ($l$). $\alpha_1$ indicates the strength of the
alternative (the larger $\alpha_1$ the stronger the
alternative). [``ordinary lrt'': classical likelihood ratio test;
``single-split'', ``multi-split'': see text; perm: permutation test as
described in text].}
\label{fig:setting1}
\end{centering}
\end{figure}

\begin{figure}[htbp!]
\begin{centering}
   \includegraphics[scale=0.65]{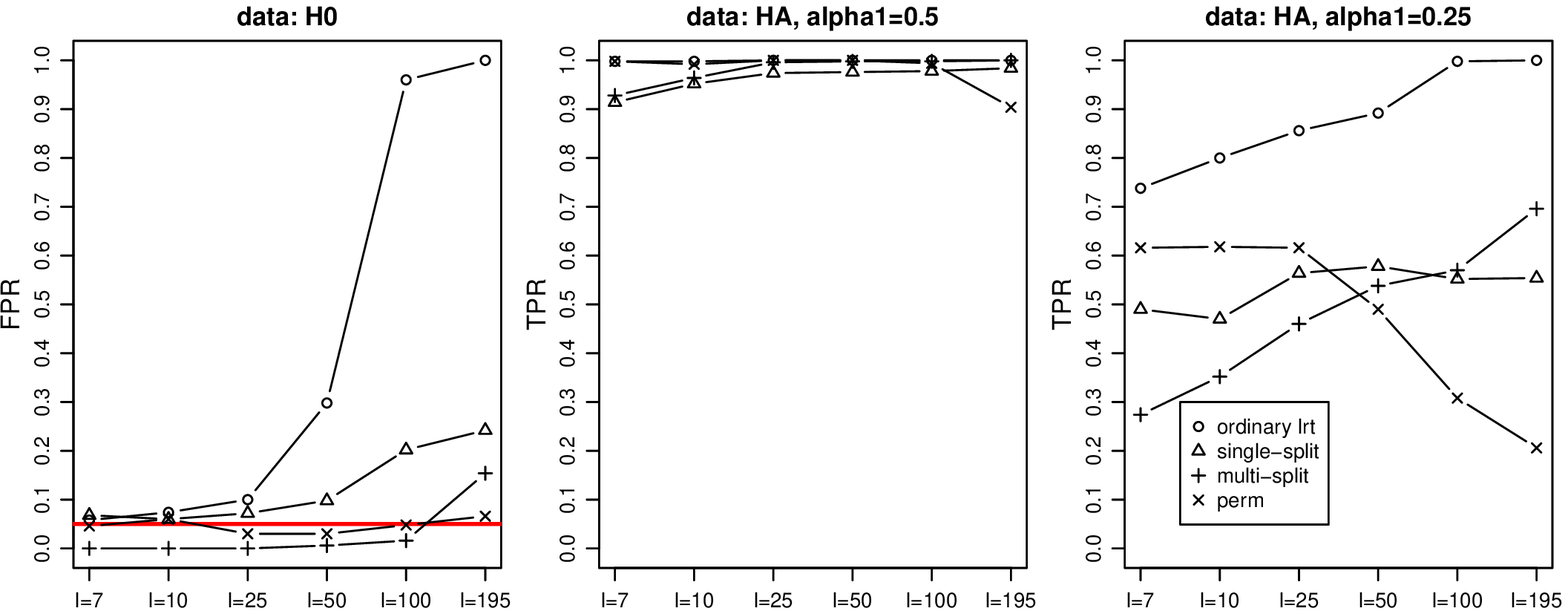} 
     \caption[]
 {Simulation study, Setting 2 (differential regression, SNR=5): False positive rate (FPR) and true positive rate
    (TPR) at 5\%-level shown as a function of number of predictors
    ($l$). $\alpha_1$ indicates the strength of the
alternative (the larger $\alpha_1$ the stronger the
alternative). [``ordinary lrt'': classical likelihood ratio test;
``single-split'', ``multi-split'': see text; perm: permutation test as
described in text].}
\label{fig:setting2}
\end{centering}
\end{figure}

\begin{figure}[htbp!]
\begin{centering}
   \includegraphics[scale=0.65]{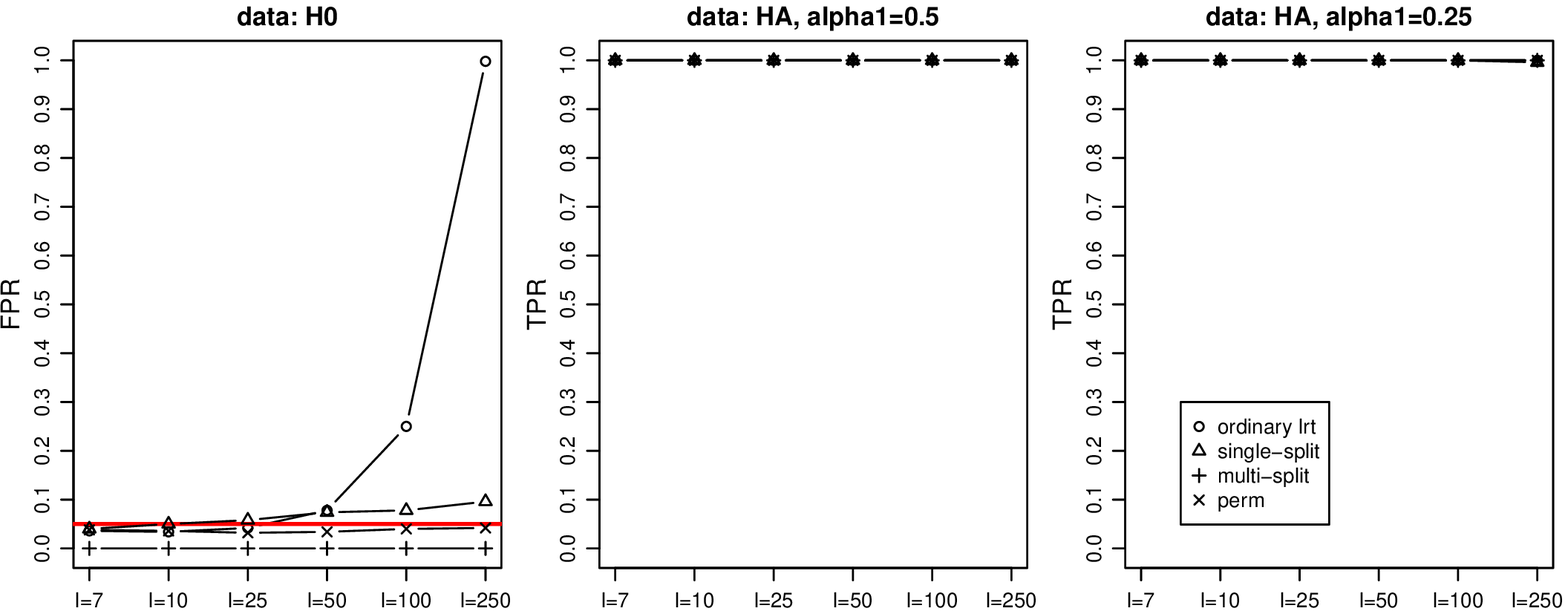} 
  \caption[]
{Simulation study, Setting 3 (differential regression,
    real predictor matrix): False positive rate (FPR) and true positive rate
    (TPR) at 5\%-level shown as a function of number of predictors
    ($l$). $\alpha_1$ indicates the strength of the
alternative (the larger $\alpha_1$ the stronger the
alternative). [``ordinary lrt'': classical likelihood ratio test;
``single-split'', ``multi-split'': see text; perm: permutation test as
described in text].}
\label{fig:setting3}
\end{centering}
\end{figure}

\begin{figure}[htbp!]
\begin{centering}
   \includegraphics[scale=0.65]{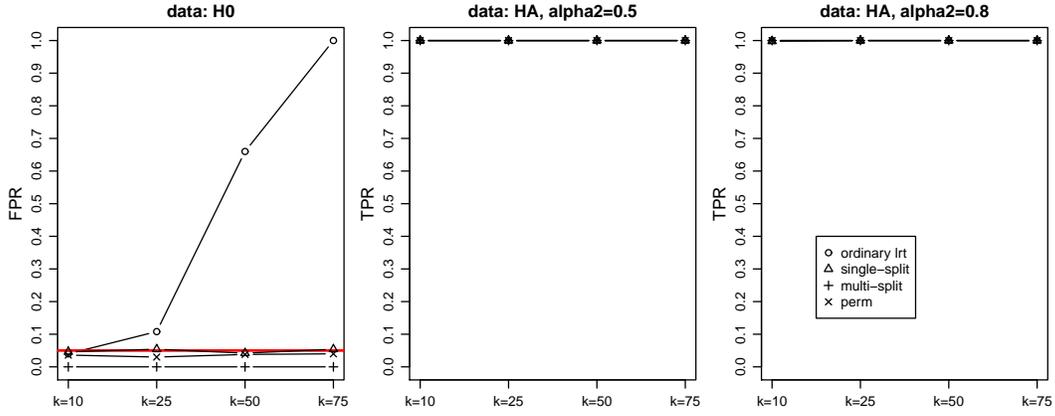} 
  \caption[]
{Simulation study, Setting 4 (differential network): False positive rate (FPR) and true positive rate
    (TPR) at 5\%-level shown as a function of number of covariates
    ($k$). $\alpha_2$ indicates the strength of the
alternative (the smaller $\alpha_1$ the stronger the
alternative). [``ordinary lrt'': classical likelihood ratio test;
``single-split'', ``multi-split'': see text; perm: permutation test as
described in text].}
\label{fig:setting4}
\end{centering}
\end{figure}

Results are shown in
Figures~\ref{fig:setting1}-\ref{fig:setting4}. The false positive rate (FPR)
of the ordinary likelihood-ratio explodes in all settings already for
medium numbers of covariates $l$, respectively $k$. We further see that the single-split method
is not able to control the true positive rate (TPR) at the 5\% level in settings
1-2. This is most probably due to the fact that the sample size is
too small and consequently the asymptotic approximation of the null distribution is not
accurate enough. The multi-split method is much more conservative and has lower FPR. This observation is in agreement with results shown in 
\cite{meinshausenpval2009}. As expected, the permutation test controls the
false positive rate at the 5\% level. Concerning the TPR (or power of the test) we find that single-splitting, multi-splitting and the
permutation test perform well in most settings. Only with weak alternatives
(setting 1 and 2 with $\alpha_1=0.25$) the TPR
decreases. Interestingly, the permutation test exhibits loss of power
for large values of $l$, whereas single- and multi-split display higher power than the permutation test 
for large $l$ but lower power for small $l$. For completeness Figures~\ref{fig:app.setting1b}-\ref{fig:app.setting4b} in
Appendix~\ref{sec:app.fig} provide more insights on the single-split
algorithm. These figures report results for settings 1-4 with respect to the sparsity and screening
properties.

\begin{figure}[htbp!]
\begin{centering}
   \includegraphics[scale=0.8]{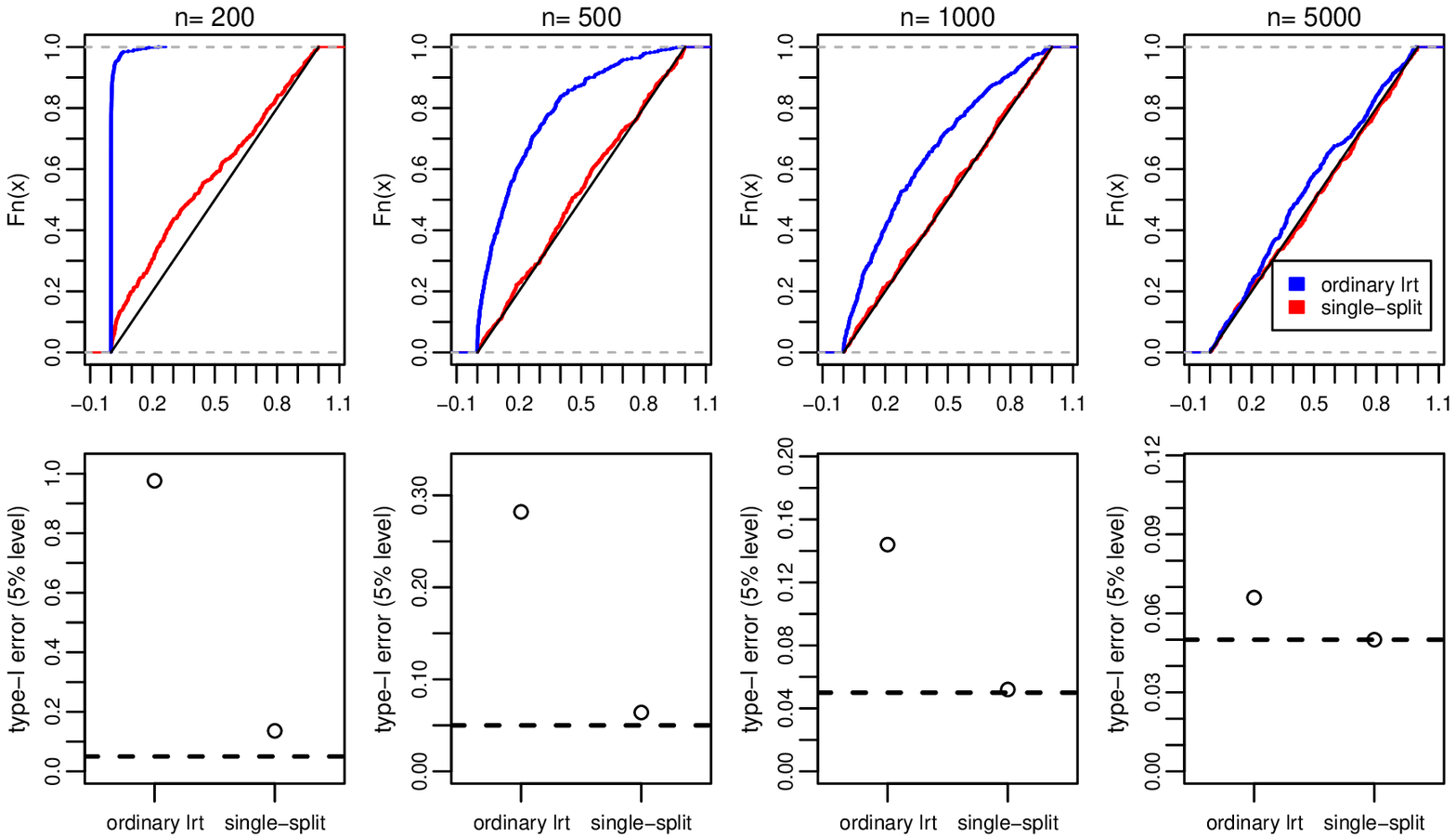} 
  \caption[]
  {Distribution of p-values, Setting 1, $l=100$. Upper row of
    panels: empirical distribution function of p-values obtained by single-split method and ordinary likelihood-ratio test for different sample sizes $(n:=)n_u=n_v=200, 500, 1000,
  5000$. Lower row of panels: fraction of p-values which are
  smaller than 0.05 (type-I error at 5\% level).  [``ordinary lrt'': classical likelihood ratio test;
``single-split'': see text].}
\label{fig:setting1c}
\end{centering}
\end{figure}

For setting 1 with $p=100$ we further examine the distribution of the
p-values obtained using the single-split method and the ordinary
likelihood-ratio test when varying the sample size
$(n:=)n_u=n_v=200, 500, 1000, 5000$. For every $n$ we use 100 samples in the
screening step of the single-split method and the rest of the samples
are used for p-value calculations. Results are shown in
Figure~\ref{fig:setting1c}. We see that p-values obtained from the single-split
method are very well approximated by the uniform distribution
function. Only in the scenario with $n=200$ the approximation is not
accurate enough which is also reflected in a too large number of false
positives.

\subsection{Application to genomic data from cancer biology}
We apply the multi-split method to real datasets from
cancer biology. For \emph{differential regression} we take data from
Broad-Novartis Cancer Cell Line Encyclopaedia (CCLE)
(\url{http://www.broadinstitute.org/ccle/home}). \cite{barretina2012}
use the data to predict anticancer drug sensitivity from genomic data. 
They describe that the histone deacetylase inhibitor \textit{panobinostat}
shows increased sensitivity in haematological cancers compared to
solid cancers.
As the response variable we take experimentally determined sensitivity to panobinostat and
we use gene expressions of the $l=100$ genes showing
highest Pearson correlation over all samples with panobinostat as the predictor
matrix. We compare the three cancer subtypes with the largest sample size:
lung cancer ($89$ cell lines), skin cancer ($71$ cell lines) and cancer
with haematopoietic and lymphoid-tissue origin (abbreviated with
\emph{haem}, $40$ cell lines). For \emph{differential
  network} we use data from The Cancer Genome
  Atlas (TCGA) data portal (\url{http://www.cancergenome.nih.gov}). We
  consider gene expression data of the $k=42$ genes present in the cancer gene list
  of \cite{weinberg2002} (downloadable at
  \url{http://www.cbio.mskcc.org/CancerGenes}) and compare $n_u=155$
  lung squamous cell carcinomas against $n_v=174$ colon adenocarcinomas. 

\begin{figure}[htbp!]
\begin{centering}
   \includegraphics[scale=0.7]{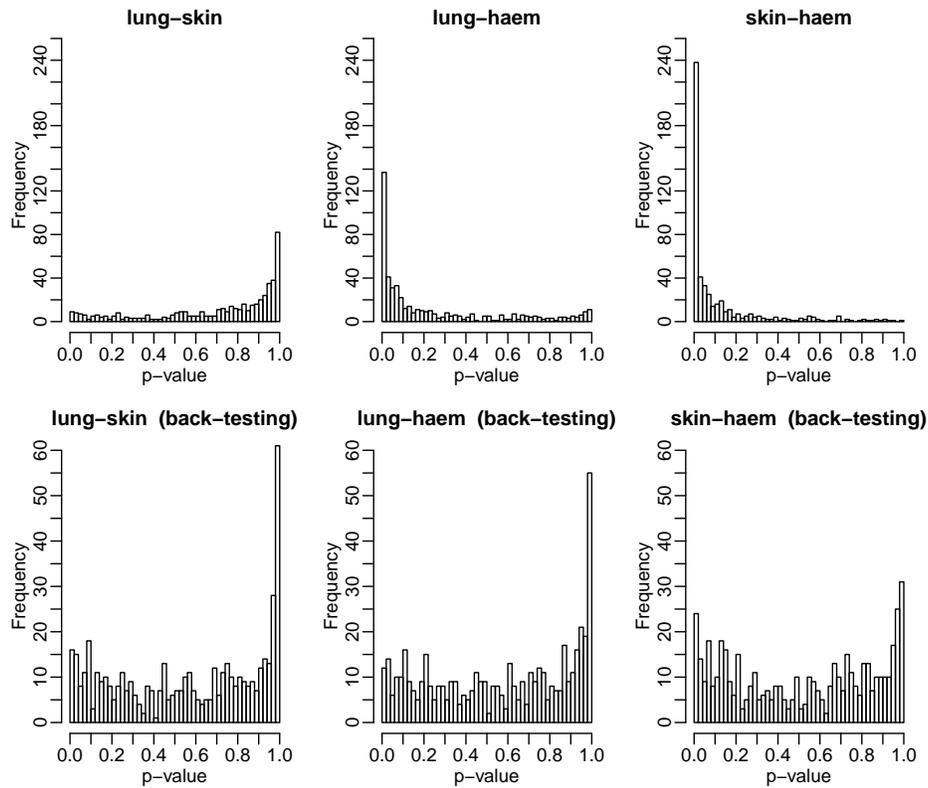} 
  \caption[]
  {Histogram p-values (Differential regression). Upper row of panels: histogram of individual p-values
    obtained from multi-split method with 500 splits comparing the different
    cancer subtypes \emph{lung}, \emph{skin} and \emph{haem}. Lower row of
    panels: histogram of p-values from ``back-testing'', where for each
    comparison the two subtypes are pooled together and then randomly
    divided into two populations.}
\label{fig:ccle}
\end{centering}
\end{figure}

For both examples we compute p-values with the
multi-split method ($500$ splits) and with the permutation test (500
permutations). In the differential regression example the sample sizes
are very small and cross-validation in the screening step can lead to
active-sets which are too large. Therefore, we adapt the
screening step by setting the smallest coefficients to zero whenever there
are more
than $\lceil 5n\rceil$ non-zero coefficients ($n$ denotes the sample
size involved in the screening step).  We further carry out
``back-testing" by dividing data randomly into
populations $U$ and $V$ and then performing significance testing. All
results are shown in Table~\ref{tab:realdata}. In the regression
example the multi-split method gives considerably
smaller p-values than the permutation test. A potential explanation could be that the permutation test exhibits small power
in difficult scenarios (see settings 1 and 2 in
Section~\ref{sec:exp.sim} with $\alpha_1=0.25$, i.e., right panel in
Figures~\ref{fig:setting1} and \ref{fig:setting2}). In the
differential network example both methods have p-values which are numerically indistinguishable from zero. For the permutation test this implies that the
observed test statistic is always larger than those obtained from
random permutations. All p-values obtained for back-testing equal one
which is reassuring. Figure~\ref{fig:ccle} shows
histograms of all $500$ p-values obtained by the multi-split
method and illustrates both the sensitivity of
p-values to single splits and also the information contained in the entire distribution of p-values obtained in the iterative, multi-split approach. For example in the
comparison of \emph{skin} against \emph{haem} we get the whole range of
p-values between zero and one. However, the distribution of these p-values is
heavily skewed towards zero which is reflected in an aggregated
p-value of 0.022.

\begin{table}[!h]
\centering
\begin{tabular}{ll|c|c|}
&&multi-split&permutation \\\hline 
Diffreg &\emph{lung-skin} &1&0.918\\
&\emph{lung-haem} &0.040&0.464\\
&\emph{skin-haem} &0.022&0.220\\
Diffnet&\emph{lusc-coad} &$<10^{-4}$&$0$\\
\end{tabular}
\caption{Genomic data examples. P-values obtained on Cancer Cell Line Encyclopedia (CCLE) and The Cancer Genome
  Atlas (TCGA). Differential
  regression (Diffreg): Anti-cancer drug panobinostat regressed on gene
  expression data; lung cancer  ($89$ samples), skin cancer ($40$
  samples) and
  haematopoietic/lymphoid cancer ($71$ samples) are compared against
  each other. Differential
  network (Diffnet): Gene expressions from $155$
  lung squamous cell carcinomas (\emph{lusc}) against $174$
  colon adenocarcinomas (\emph{coad}).}
\label{tab:realdata}
\end{table}

\section{Discussion and conclusions}\label{sec:discussion}
We have presented a novel and very general approach for high-dimensional two-sample
testing. We combined ideas including sample-splitting,
non-nested hypothesis testing and p-value aggregation to propose a methodology that allows 
two-sample  testing 
for a wide range of
high-dimensional models. We treated in detail linear regression
(\emph{differential regression}) and
Gaussian graphical models (\emph{differential network}) and validated
their performance on simulated and real data.

Our methodology is supported by asymptotic theory. However, our
results obtained in Proposition~\ref{prop:lrtHA} and
Theorem~\ref{thm:asy.null} do not reflect a truly high-dimensional
setup as they consider the active sets selected in the
screening step as fixed. An aim of our ongoing research is to generalize these results
to the case where the size of the active sets can grow with a rate of
smaller order than the sample size.

Whilst the focus of this paper was not on applications, we 
note that the methodology we propose should have immediate utility in biology. 
Differential network can be used to test a number of hypotheses of current scientific interest.
For example, in cancer biology it is widely believed that genomic differences between cancer types and subtypes
may in some cases be manifested also at the level of biological networks, including gene regulatory  and protein signaling networks.
However, we are not aware of existing methodology that allows such hypotheses to be tested 
statistically under the relevant conditions of moderate sample size and high dimensionality.
Our approach can be used to test such hypotheses directly from high-throughput data, as illustrated in the example above.
A further application of the differential networks formulation of our approach is in gene-set testing.
Currently, gene set tests (e.g., \cite{subramanian2005,irizarry2009}) are not truly multivariate. Our approach could be directly applied to 
test differences in gene sets at the level of not only means but also covariances or networks.

 
 \vspace{1cm}
 
 \noindent
\textbf{Acknowledgements:} The authors thank Peter B\"{u}hlmann for discussions. This work was supported in part by NCI U54 CA 112970 and the
Cancer Systems Biology Center grant from the Netherlands Organisation for Scientific Research.

\appendix

\section{Proofs of Section~\ref{sec:method}}\label{sec:app.proof}
Without loss of generality, we take $(n:=)=n_u=n_v$.
We assume that the following assumptions (A1)-(A6) hold:

\begin{itemize}
\item[(A1)] Data $\U=(\Y_u,\X_u)$ and $\V=(\Y_v,\X_v)$ are
  i.i.d. samples with $X_u, X_v$ drawn from some
common density function and $Y_u|X_u=x\sim d(\cdot|x;\phi_u)$, $Y_v|X_v=x\sim d(\cdot|x;\phi_v)$. The conditional density function $d(y|x;\phi)$ is strictly
positive for almost all $(y,x)$ and all~$\phi$. 


\item[(A2)] $\Phi_u$, $\Phi_v$ and $\Phi_{uv}$ are compact subsets of $\R^p$,
and the conditional density function $d(y|x;\phi)$ is continuous in $\phi$.

\item[(A3)]  For almost all $(y,x)$, $|\log d(y|x;\cdot)|$ is dominated by an integrable function
independent of $\phi$. The functions
\begin{eqnarray*}
 \phi_{u} &\mapsto& \E[\E_{\phi_u}[\log
  d(Y_u|X_u;\phi_{u})]]\\
  \phi_{v} &\mapsto& \E[\E_{\phi_v}[\log
  d(Y_v|X_v;\phi_{v})]]\\
\phi_{uv} &\mapsto& \E[\E_{\phi_u}[\log
  d(Y_u|X_u;\phi_{uv})]]+\E[\E_{\phi_v}[\log
  d(Y_v|X_v;\phi_{uv})]]
\end{eqnarray*}
have unique maximums on $\Phi_u$, $\Phi_v$ and $\Phi_{uv}$ at
$\phi^*_{u}$, $\phi^*_{v}$ and $\phi^*_{uv}$.

\item[(A4)] For almost all $(y,x)$: $\log d(y|x;\cdot)$ is twice continuously differentiable on
$\Phi$;  $|\partial\log d(y|x;\cdot)/\partial \phi \cdot \partial\log
d(y|x;\cdot)/\partial \phi' |$ and $|\partial\log
d(y|x;\cdot)/\partial\phi\partial\phi'|$ are dominated by integrable
functions independent on $\phi$.

\item[(A5)] If $(\bar{\phi}:=)\phi_u=\phi_v$, then $\bar{\phi}$ is an interior point of
$\Phi_u$, $\Phi_v$ and $\Phi_{uv}$. Further, $\bar{\phi}$ is a regular
point of $B_{M_{\rm ind}}$ and $B_{M_{\rm joint}}$. 

\item[(A6)] The information matrix equivalence holds, i.e., for all
  $A\subset\{1,\ldots,p\}$ we have:
$$\E[\E_{\bar{\phi}}[\partial\log
d(y|x;\bar{\phi})/\partial\phi_{A}\cdot\partial\log
d(y|x;\bar{\phi})/\partial\phi_{A}]]=-\E[\E_{\bar{\phi}}[\partial^2\log
d(y|x;\bar{\phi})/\partial\phi_{A}^2]].$$

\end{itemize}

{\bf Proof of Proposition~\ref{prop:lrtHA}}
Given assumptions (A1)-(A3) and noting that $\phi^*_u=\phi_u$,
$\phi^*_v=\phi_v$ (a consequence of the screening property) we have
\begin{eqnarray}\label{eq:proplrtHA}
&\frac{1}{n}\mathrm{LR_{n,n}}\overset{a.s.}{\rightarrow}\E_{(\phi_u,\phi_v)}\!\left[\log\frac{d(Y_u|X_u;\phi_u)d(Y_v|X_v;\phi_v)}{d(Y_u|X_u;\phi^*_{uv})d(Y_v|X_v;\phi^*_{uv})}\right]=\D\left(\phi_u\|\phi^*_{uv}\right)+\D\left(\phi_v\|\phi^*_{uv}\right)\!,
\end{eqnarray}
where
$\D(\phi\|\phi')=\E\left[\E_{\phi}\left[\log\frac{d(Y|X;\phi)}{d(Y|X;\phi')}\right]\right]$
is the standard Kullback-Leibler divergence.
If $\phi_u\neq\phi_v$, then at least one term on the right hand side
of equation~(\ref{eq:proplrtHA}) is strictly positive. Therefore,
under $\mathbf{H_A}$: $$\mathrm{LR_{n,n}}\overset{a.s.}{\rightarrow}\infty \quad (n\rightarrow\infty).$$
\hfill$\Box$
\vskip 0.2in

{\bf Proof of Theorem~\ref{thm:asy.null}}
Consider the setting in \cite{vuong1989} with competing models $\mathbf{F_{\phi}}=M_{\rm ind}$ and $\mathbf{G_{\gamma}}=M_{\rm joint}$, i.e.,
\begin{eqnarray*}
&f(y|x;\theta)=d(y_u|x_u;\phi_u)d(y_v|x_v;\phi_v)\quad\mathrm{and}\quad
g(y|x;\gamma)=d(y_u|x_u;\phi_{uv})d(y_v|x_v;\phi_{uv})
\end{eqnarray*}
with $y=(y_u,y_v)$, $x=(x_u,x_v)$, $\theta=(\phi_u,\phi_v)$ and
$\gamma=\phi_{uv}$. 
If the screening property holds then model $M_{\rm ind}$ is correctly specified and
the pseudo-true values $(\phi_u^*,\phi_v^*)$ equal the true values
$(\phi_u,\phi_v)$. Furthermore, if we have $(\bar{\phi}:=)\phi_u=\phi_v$ then the
screening property guarantees that also $M_{\rm joint}$ is correctly
specified and that $\phi_{uv}^*=\bar{\phi}$. As a consequence we have
$f(y|x;\theta^*)=g(y|x;\gamma^*)$. Therefore, assuming assumptions
(A1)-(A5), it follows from Theorem~3.3 (i) in \cite{vuong1989}
that
$$\textrm{LR}_{n,n}\overset{d}{\rightarrow} \Psi_r(\cdot;\nu),$$ where
$r=|I_u|+|I_v|+|I_{uv}|$ and the weights
$\nu_j,\;j=1,\ldots,r,$ are
eigenvalues of the matrix 
\begin{eqnarray}
W&=&\begin{pmatrix}
  -B_f  A_f^{-1}& -B_{fg}A_g^{-1}\\
   B_{gf}A_f^{-1}&  B_g  A_g^{-1}\\
\end{pmatrix}.
\end{eqnarray}
The matrices $B_f$, $B_g$, $B_{fg}$, $B_{gf}$, $A_f$ and $A_g$ are defined as
in \cite{vuong1989}. If we set $$B_{M_{\rm
    ind}}=B_f,\; B_{M_{\rm
    joint}}=B_g\; \textrm{and}\; B_{M_{\rm joint}M_{\rm
    ind}}=B_{gf}$$ then we obtain
as a consequence of the independence of the populations $U$ and $V$:
\begin{eqnarray*}
&B_{M_{\rm ind}}=\begin{pmatrix}
  B^u_{I_u}& 0\\
  0&  B^v_{I_v}\\
\end{pmatrix},\quad B_{M_{\rm joint}}=B^u_{I_{uv}}+B^v_{I_{uv}}
\end{eqnarray*}
and $$B_{M_{\rm joint} M_{\rm
    ind}}=\left(B^u_{I_{uv}I_u},B^v_{I_{uv}I_v}\right).$$ 
By invoking the information matrix equivalence, assumption (A6), we
find
\begin{eqnarray*}
W&=&\begin{pmatrix}
  \ID_{r_u+r_v}& B_{M_{\rm ind} M_{\rm joint}} B^{-1}_{M_{\rm joint}}\\
   B_{M_{\rm joint} M_{\rm ind}}B^{-1}_{M_{\rm ind}}&  -\ID_{r_{uv}}\\

 \end{pmatrix}.
\end{eqnarray*}

\hfill$\Box$
\vskip 0.2in

{\bf Proof of Proposition~\ref{prop:evals}}
Set $r_u=|I_u|$, $r_v=|I_v|$, $r_{uv}=|I_{uv}|$ and
$r=r_u+r_v+r_{uv}.$ The eigenvalues of $W$ are the solutions to the equation

\begin{eqnarray*}
  \det(W-\nu\ID)&=\det\begin{pmatrix}
  \ID_{r_u+r_v}-\nu& B_{M_{\rm ind} M_{\rm joint}} B^{-1}_{M_{\rm joint}}\\
   B_{M_{\rm joint} M_{\rm ind}}B^{-1}_{M_{\rm ind}}&  -\ID_{r_{uv}}-\nu\\
\end{pmatrix}=0.
\end{eqnarray*}

Assume $|I_u|+|I_v|\geq |I_{uv}|.$ 

If $\nu\neq 1$, then we find by using Schur's complement
\begin{eqnarray}\label{eq:app0}
  &\det(W-\nu\ID)\!=\!\det(\ID_{r_u\! +\!r_v}\!-\!\nu)\!\det(B_{M_{\rm joint} M_{\rm ind}}B^{-1}_{M_{\rm ind}}B_{M_{\rm ind}
    M_{\rm joint}}B^{-1}_{M_{\rm joint}}\!-\!\mu\ID_{r_{uv}}),
\end{eqnarray}
where $\mu=(1-\nu^2)$. The second term on the right of
equation~(\ref{eq:app0}) has $r_{uv}$ roots. Therefore $r_u
+r_v-r_{uv}$ of the eigenvalues
$\nu_j$ equal one. As $\phi^*_u=\phi^*_v=\phi^*_{uv}$, we can write $$B^u_{I_{u} I_{uv}}=B^u_{I_{u}
  I_{uv}}(\phi^*_{u};\phi^*_{uv})\quad \textrm{and} \quad B^v_{I_{v} I_{uv}}=B^v_{I_{v}
  I_{uv}}(\phi^*_{v};\phi^*_{uv}). $$
Solving
\begin{eqnarray*}
 & \det(B_{M_{\rm joint} M_{\rm ind}}B^{-1}_{M_{\rm ind}}B_{M_{\rm ind}
    M_{\rm joint}}B^{-1}_{M_{\rm joint}}-\mu\ID_{r_{uv}})=0
\end{eqnarray*}
is equivalent to 
\begin{eqnarray}\label{eq:app1}
&  \det\left((B^u_{I_{uv}I_u}(B^u_{I_u})^{-1}B^u_{I_u I_{uv}}-\mu B^u_{I_{uv}})+(B^v_{I_{uv}I_v}(B^v_{I_v})^{-1}B^v_{I_v
    I_{uv}}-\mu B^v_{I_{uv}})\right)=0.
\end{eqnarray}
An easy calculation involving Schur's complement shows
\begin{eqnarray*}
  (B^u_{I_u})^{-1}&=\begin{pmatrix}
 B^u_J& B^u_{J\minI_u}\\
   B^u_{\minI_u J}&B^u_{\minI_u}\\

 \end{pmatrix}^{-1}\!=\!\begin{pmatrix}
 (B^u_J)^{-1}\!+\!(B^u_J)\!^{-1}B^u_{J\minI_u}\!(Q^u_{\minI_u})^{-1}\!B^u_{\minI_uJ}\!(B^u_J)^{-1}&-(B^u_J)^{-1}\!B^u_{J\minI_u}\!(Q^u_{\minI_u})^{-1}\\
   -(Q^u_{\minI_u})^{-1}B^u_{\minI_uJ}(B^u_J)^{-1}&(Q^u_{\minI_u})^{-1}\\
\end{pmatrix}.
\end{eqnarray*}
Further, we find
\begin{eqnarray*}
   B^u_{I_{uv}I_u}(B^u_{I_u})^{-1}B^u_{I_u I_{uv}}&=&\begin{pmatrix}
 B^u_J& B^u_{J\minI_{uv}}\\
   B^u_{\minI_{uv} J}&Q^u_{\minI_{uv}\minI_u}(Q^{u}_{\minI_u})^{-1}Q^u_{\minI_u \minI_{uv}}+B^u_{\minI_{uv} J}(B^{u}_{J})^{-1}B^u_{J\minI_{uv}}\\

 \end{pmatrix}. \end{eqnarray*}
Putting together, equation~(\ref{eq:app1}) reads
\begin{eqnarray}
&0=\det\begin{pmatrix}\label{eq:app2}
 (1\!-\!\mu)(B^u_J\!+\!B^v_J)& (1-\mu)(B^u_{J\minI_{uv}}\!+\!B^v_{J\minI_{uv}})\\
   (1\!-\!\mu)(B^u_{\minI_{uv} J}\!+\!B^v_{\minI_{uv} J})&*\\

 \end{pmatrix}
\end{eqnarray}
with \begin{eqnarray*}*&=&\left(Q^u_{\minI_{uv}\minI_u}(Q^{u}_{\minI_u})\!^{-1}Q^u_{\minI_u
     \minI_{uv}}+Q^v_{\minI_{uv}\minI_v}(Q^{v}_{\minI_v})\!^{-1}Q^v_{\minI_v
     \minI_{uv}}\right)\\&&+B^u_{\minI_{uv}
   J}(B^{u}_{J})\!^{-1}B^u_{J\minI_{uv}}+B^v_{\minI_{uv}
   J}(B^{v}_{J})\!^{-1}B^v_{J\minI_{uv}}-\mu
 (B^u_{\minI_{uv}}+B^v_{\minI_{uv}}).\\
\end{eqnarray*}
From (\ref{eq:app2}) we deduce
\begin{eqnarray*}
0&=&(1-\mu)^{|J|}\left(*-(1-\mu)(B^u_{\minI_{uv} J}+B^v_{\minI_{uv} J})(B^u_{J}+B^v_{J})^{-1}(B^u_{J \minI_{uv}}+B^v_{J \minI_{uv}})\right)\\
&=&(1-\mu)^{|J|}\det\left(\Big(Q^u_{\minI_{uv}\minI_u}(Q^{u}_{\minI_u})^{-1}Q^u_{\minI_u
     \minI_{uv}}+Q^v_{\minI_{uv}\minI_v}(Q^{v}_{\minI_v})^{-1}Q^v_{\minI_v
     \minI_{uv}}\Big)(Q^u_{\minI_{uv}}+Q^{v}_{\minI_{uv}})^{-1}-\mu\ID_{r_{uv}}\right)\\
 \end{eqnarray*}
and conclude that $2|J|$ of the eigenvalues are zero and that the
remaining eigenvalues are obtained by computing eigenvalues
of  $$\big(Q_{\minI_{uv}\minI_u}Q^{-1}_{\minI_u}Q_{\minI_u\minI_{uv}}+Q_{\minI_{uv}\minI_v}Q^{-1}_{\minI_v}Q_{\minI_v\minI_{uv}}\big)(2Q_{\minI_{uv}})^{-1}.$$
The proof for the case with $|I_{uv}|>|I_u|+|I_v|$ works similarly.
\hfill$\Box$
\vskip 0.2in
\section{Derivation of equations (\ref{eq:betamat.regr}) and (\ref{eq:betamat.ggm}) in Section~\ref{sec:regr.ggm}}
\paragraph{Equation (\ref{eq:betamat.regr})}\label{sec:app.expect}
Set $\epsilon_c=Y-X^{\top}\beta_c$ and note that
$\E_{\phi_c}[\epsilon_c]=0,\;\E_{\phi_c}[\epsilon_c^2]=\sigma_c^2$.
We then find:
\begin{eqnarray*}
\E_{\phi_c}\left[s(Y|X;\phi_a)s(Y|X;\phi_b)^T\right]&=&\E_{\phi_c}\left[\left(\frac{(Y-X^T\beta_a)X}{\sigma_a^2}\right)\left(\frac{(Y-X^T\beta_b)X}{\sigma_b^2}\right)^T\right]\\
&=&\frac{XX^T}{\sigma_a^2\sigma_b^2}\E_{\phi_c}\left[(\epsilon_c+X^{T}(\beta_c-\beta_a))(\epsilon_c+X^T(\beta_c-\beta_b))\right]\\
&=&\frac{XX^T}{\sigma_a^2\sigma_b^2}\E_{\phi_c}\left[\epsilon_c^2\!+\!\epsilon_c X^T(\beta_c\!-\!(\beta_a\!+\!\beta_b))\!+\!(\beta_c\!-\!\beta_a)^TXX^T(\beta_c\!-\!\beta_b)\right]\\
&=&\frac{XX^T}{\sigma_a^2\sigma_b^2}(\sigma_c^2+(\beta_c-\beta_a)^T XX^T (\beta_c-\beta_b)).
\end{eqnarray*}
\paragraph{Equation (\ref{eq:betamat.ggm})} 
Invoking formulas on fourth moments of a multivariate Normal
distribution we find:
\begin{eqnarray*}
\E_{\phi_c}\left[s_{(j,j')}(Y;\phi_a)s_{(l,l')}(Y;\phi_b)\right]&=&\E_{\phi_c}\left[(Y^{(j)}Y^{(j')}-\Sigma_{jj'})(Y^{(l)}Y^{(l')}-\Sigma_{ll'})\right]\\
&=&\E_{\phi_c}\left[Y^{(j)}Y^{(j')}Y^{(l)}Y^{(l')}\!-\!Y^{(j)}Y^{(j')}\Sigma_{b,ll'}\!-\!Y^{(l)}Y^{(l')}\Sigma_{a,jj'}\!+\!\Sigma_{a,jj'}\Sigma_{b,ll'}\right]\\
&=&\Sigma_{c,jj'}\Sigma_{c,ll'}+\Sigma_{c,jl}\Sigma_{c,j'l'}+\Sigma_{c,jl'}\Sigma_{c,j'l}\\
&&-\Sigma_{c,jj'}\Sigma_{b,ll'}-\Sigma_{c,ll'}\Sigma_{a,jj'}+\Sigma_{a,jj'}\Sigma_{b,ll'}.
\end{eqnarray*}
\section{Permutation test and symmetric Kullback-Leibler divergence}\label{sec:app.perm}
The symmetric Kullback Leibler divergence is defined as 
\begin{eqnarray*}
\D_{\rm symm}(\phi\|\phi')&=&\D(\phi\|\phi')+\D(\phi'\|\phi).
\end{eqnarray*}
Consider a random partitition of the data into two groups $\tilde{\U}=(\Y_{\tilde u},\X_{\tilde u})$ and $\tilde{\V}=(\Y_{\tilde v},\X_{\tilde v})$. We construct a permutation test based on the test
statistic $$\D_{\rm symm}(\hat\phi_{\tilde{u},\lambda_{\rm
    cv}}\|\hat\phi_{\tilde{v},\lambda_{\rm cv}}),$$
with
$$\hat{\phi}_{\tilde{u},\lambda}=\argmax\limits_{\phi\in\Phi}\ell(\phi;\Y_{\tilde u}|\X_{\tilde u})-\lambda\|\phi\|_1,\quad\hat{\phi}_{\tilde{v},\lambda}=\argmax\limits_{\phi\in\Phi}\ell(\phi;\Y_{\tilde v}|\X_{\tilde v})-\lambda\|\phi\|_1$$
and tuning parameter chosen by
cross-validation ($\lambda=\lambda_{\rm cv}$).

For the regression model the
Kullback-Leibler divergence equals:
\begin{eqnarray*}
\D\left((\beta,\sigma^2)\|(\beta',\sigma'^2)\right)&=&-\frac{1}{2}+\frac{\sigma^2}{2\sigma'^2}+(\beta-\beta')^T\E[xx^T](\beta-\beta')+\frac{1}{2}\log\frac{\sigma'^2}{\sigma^2}.
\end{eqnarray*}
For the Gaussian graphical model we have: 
\begin{eqnarray*}
    \D(\Omega\|\Omega')&=& {1 \over 2} \left( \mathrm{tr} \left(\Omega'\Omega^{-1} \right) -\log \left( { \det\Omega \over \det\Omega' } \right) - k\right).
\end{eqnarray*} 
\newpage
\section{Additional Figures of Section~\ref{sec:exp.num}}\label{sec:app.fig}
\begin{figure}[htbp!]
\begin{centering}
   \includegraphics[scale=0.5]{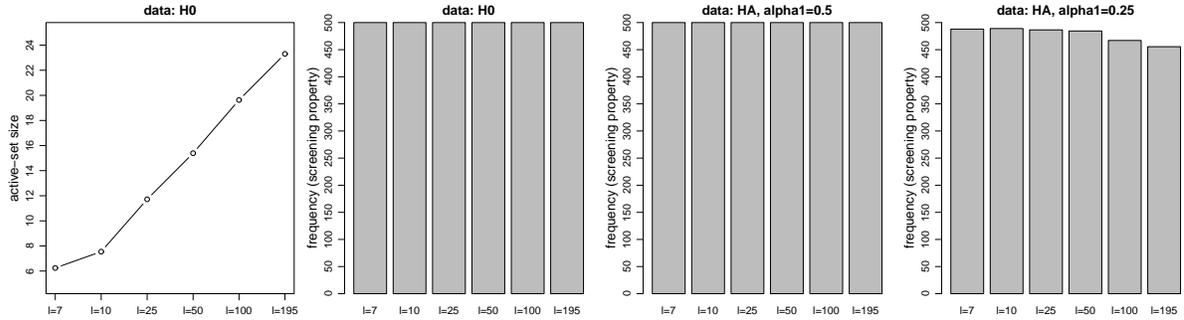} 
  \caption[]
  {Simulation study, Setting 1 (differential regression, SNR=10):
    Sparsity and screening property of single-split method over 500 simulation
    runs. Left panel shows the
    average number of non-zero parameter components of the joint model selected
    in the screening step (i.e., $|I_{uv}|$) . The second to the fourth panels show
    frequencies which indicate how many times the screening property is satisfied.
    }
\label{fig:app.setting1b}
\end{centering}
\end{figure}
\begin{figure}[htbp!]
\begin{centering}
\includegraphics[scale=0.5]{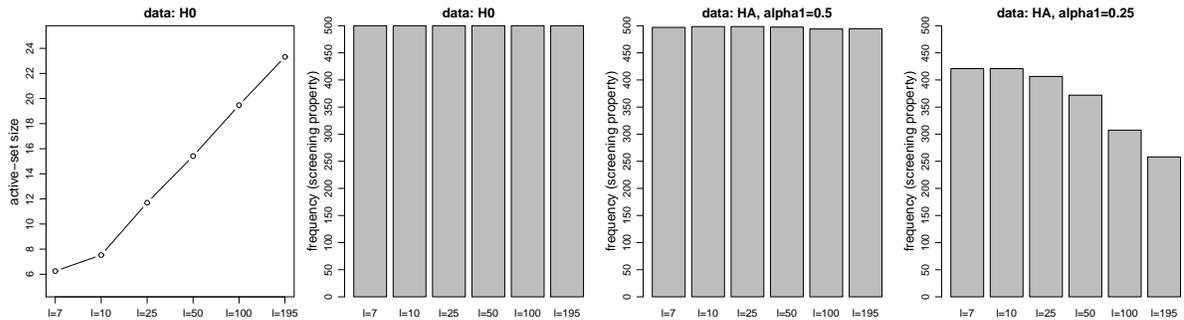} 
  \caption[]
  {Simulation study, Setting 2 (differential regression, SNR=5): Same notation as in Figure~\ref{fig:app.setting1b}.}
\label{fig:app.setting2b}
\end{centering}
\end{figure}
\begin{figure}[htbp!]
\begin{centering}
\includegraphics[scale=0.5]{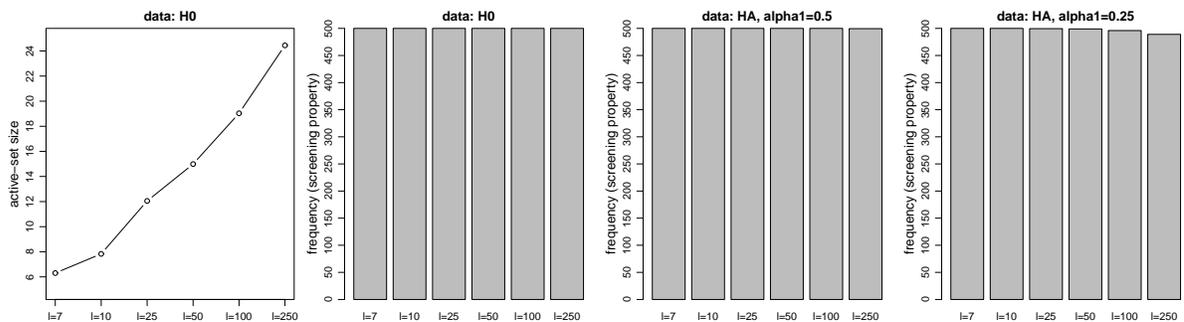} 
\caption[]
  {Simulation study, Setting 3 (differential regression, real predictor matrix):
    Same notation as in Figure~\ref{fig:app.setting1b}.}
\label{fig:app.setting3b}
\end{centering}
\end{figure}
\begin{figure}[htbp!]
\begin{centering}
\includegraphics[scale=0.5]{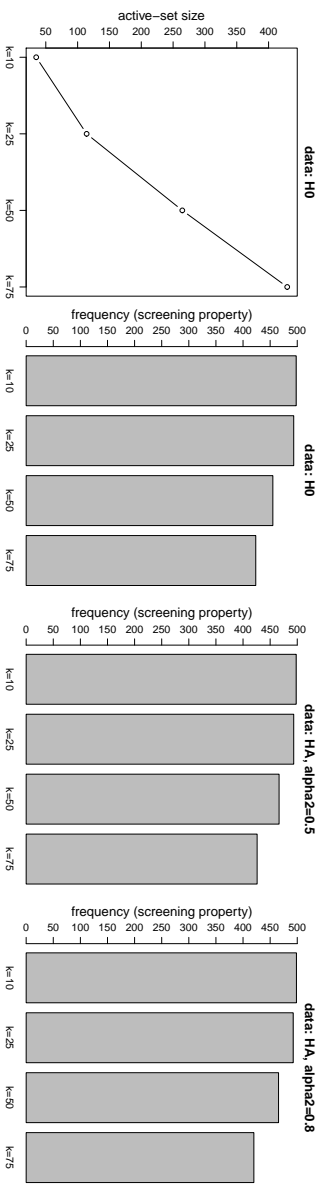} 
  \caption[]
  {Simulation study, Setting 4 (differential network): Same notation as in Figure~\ref{fig:app.setting1b}.}
\label{fig:app.setting4b}
\end{centering}
\end{figure}





\begin{thebibliography}{24}
\expandafter\ifx\csname natexlab\endcsname\relax\def\natexlab#1{#1}\fi
\expandafter\ifx\csname url\endcsname\relax
  \def\url#1{\texttt{#1}}\fi
\expandafter\ifx\csname urlprefix\endcsname\relax\def\urlprefix{URL }\fi

\bibitem[{Bai and Saranadasa(1996)}]{bai1996}
Bai, Z. and Saranadasa, H. (1996) Effect of high dimension: By an example of a
  two sample problem.
\newblock \emph{Statistica Sinica}, \textbf{6}, 311--329.

\bibitem[{Barretina \emph{et~al.}(2012)Barretina, Caponigro, Stransky,
  Venkatesan, Margolin, Kim, Wilson, Leh\'{a}r, Kryukov, Sonkin, Reddy, Liu,
  Murray, Berger, Monahan, Morais, Meltzer, Korejwa, Jan\'{e}-Valbuena, Mapa,
  Thibault, Bric-Furlong, Raman, Shipway, Engels, Cheng, Yu, Yu, Aspesi,
  de~Silva, Jagtap, Jones, Wang, Hatton, Palescandolo, Gupta, Mahan, Sougnez,
  Onofrio, Liefeld, MacConaill, Winckler, Reich, Li, Mesirov, Gabriel, Getz,
  Ardlie, Chan, Myer, Weber, Porter, Warmuth, Finan, Harris, Meyerson, Golub,
  Morrissey, Sellers, Schlegel and Garraway}]{barretina2012}
Barretina, J., Caponigro, G., Stransky, N., Venkatesan, K., Margolin, A.~A.,
  Kim, S., Wilson, C.~J., Leh\'{a}r, J., Kryukov, G.~V., Sonkin, D., Reddy, A.,
  Liu, M., Murray, L., Berger, M.~F., Monahan, J.~E., Morais, P., Meltzer, J.,
  Korejwa, A., Jan\'{e}-Valbuena, J., Mapa, F.~A., Thibault, J., Bric-Furlong,
  E., Raman, P., Shipway, A., Engels, I.~H., Cheng, J., Yu, G.~K., Yu, J.,
  Aspesi, P., de~Silva, M., Jagtap, K., Jones, M.~D., Wang, L., Hatton, C.,
  Palescandolo, E., Gupta, S., Mahan, S., Sougnez, C., Onofrio, R.~C., Liefeld,
  T., MacConaill, L., Winckler, W., Reich, M., Li, N., Mesirov, J.~P., Gabriel,
  S.~B., Getz, G., Ardlie, K., Chan, V., Myer, V.~E., Weber, B.~L., Porter, J.,
  Warmuth, M., Finan, P., Harris, J.~L., Meyerson, M., Golub, T.~R., Morrissey,
  M.~P., Sellers, W.~R., Schlegel, R. and Garraway, L.~A. (2012) {The Cancer
  Cell Line Encyclopedia enables predictive modelling of anticancer drug
  sensitivity.}
\newblock \emph{Nature}, \textbf{483}, 603--607.

\bibitem[{B\"uhlmann(2012)}]{buhlmann2012}
B\"uhlmann, P. (2012) Statistical significance in high-dimensional linear
  models.
\newblock \emph{arXiv.org: 1202.1377}.

\bibitem[{B\"uhlmann and van~de Geer(2011)}]{lassobook2011}
B\"uhlmann, P. and van~de Geer, S. (2011) \emph{Statistics for High-Dimensional
  Data: Methods, Theory and Applications}.
\newblock Springer Series in Statistics. Springer.

\bibitem[{Cai \emph{et~al.}(2011)Cai, Liu and Xia}]{tonycai2011}
Cai, T., Liu, W. and Xia, Y. (2011) Two-sample covariance matrix testing and
  support recovery in high-dimensional and sparse settings.
\newblock Tech. rep., Department of Statistics, The Wharton School, University
  of Pennsylvania.

\bibitem[{Chen and Qin(2010)}]{chen2010}
Chen, S.~X. and Qin, Y.-L. (2010) A two-sample test for high-dimensional data
  with applications to gene-set testing.
\newblock \emph{Annals of Statistics}, \textbf{38}, 808--835.

\bibitem[{Davies(1980)}]{davies1980}
Davies, R.~B. (1980) Algorithm {AS} 155: The distribution of a linear
  combination of chi-2 random variables.
\newblock \emph{Journal of the Royal Statistical Society. Series C},
  \textbf{29}, 323--333.

\bibitem[{Duchesne and de~Micheaux(2010)}]{duchesne2012}
Duchesne, P. and de~Micheaux, P.~L. (2010) Computing the distribution of
  quadratic forms: further comparisons between the {L}iu-{T}ang-{Z}hang
  approximation and exact methods.
\newblock \emph{Computational Statistics and Data Analysis}, \textbf{54},
  858--862.

\bibitem[{Fan and Li(2001)}]{fan01variable}
Fan, J. and Li, R. (2001) Variable selection via penalized likelihood.
\newblock \emph{Journal of the American Statistical Association}, \textbf{96},
  1348--1360.

\bibitem[{Friedman \emph{et~al.}(2008)Friedman, Hastie and
  Tibshirani}]{friedman2007sic}
Friedman, J., Hastie, T. and Tibshirani, R. (2008) Sparse inverse covariance
  estimation with the graphical {L}asso.
\newblock \emph{Biostatistics}, \textbf{9}, 432--441.

\bibitem[{Friedman \emph{et~al.}(2010)Friedman, Hastie and
  Tibshirani}]{friedman2010}
Friedman, J., Hastie, T. and Tibshirani, R. (2010) Regularization paths for
  generalized linear models via coordinate descent.
\newblock \emph{Journal of Statistical Software}, \textbf{33}, 1--22.

\bibitem[{Hahn and Weinberg(2002)}]{weinberg2002}
Hahn, W.~C. and Weinberg, R.~A. (2002) Modelling the molecular circuitry of
  cancer.
\newblock \emph{Nature Reviews Cancer}, \textbf{2}, 331--341.

\bibitem[{Irizarry \emph{et~al.}(2009)Irizarry, Wang, Zhou and
  Speed}]{irizarry2009}
Irizarry, R.~A., Wang, C., Zhou, Y. and Speed, T.~P. (2009) {Gene set
  enrichment analysis made simple}.
\newblock \emph{Statistical Methods in Medical Research}, \textbf{18},
  565--575.

\bibitem[{Li and Chen(2012)}]{li2012}
Li, J. and Chen, S.~X. (2012) Two sample tests for high-dimensional covariance
  matrices.
\newblock \emph{Annals of Statistics}, \textbf{40}, 908--940.

\bibitem[{Lopes \emph{et~al.}(2012)Lopes, Jacob and Wainwright}]{lopes2012}
Lopes, M., Jacob, L. and Wainwright, M.~J. (2012) A more powerful two-sample
  test in high dimensions using random projection.
\newblock \emph{arXiv.org:1108.2401}.

\bibitem[{Meinshausen and B\"uhlmann(2006)}]{meinshausen04consistent}
Meinshausen, N. and B\"uhlmann, P. (2006) High dimensional graphs and variable
  selection with the {L}asso.
\newblock \emph{Annals of Statistics}, \textbf{34}, 1436--1462.

\bibitem[{Meinshausen \emph{et~al.}(2009)Meinshausen, Meier and
  B\"uhlmann}]{meinshausenpval2009}
Meinshausen, N., Meier, L. and B\"uhlmann, P. (2009) P-values for
  high-dimensional regression.
\newblock \emph{Journal of the American Statistical Association}, \textbf{104},
  1671--1681.

\bibitem[{Park and Hastie(2007)}]{Park2007}
Park, M.~Y. and Hastie, T. (2007) {L1-regularization path algorithm for
  generalized linear models}.
\newblock \emph{Journal of the Royal Statistical Society: Series B},
  \textbf{69}, 659--677.

\bibitem[{Subramanian \emph{et~al.}(2005)Subramanian, Tamayo, Mootha,
  Mukherjee, Ebert, Gillette, Paulovich, Pomeroy, Golub, Lander and
  Mesirov}]{subramanian2005}
Subramanian, A., Tamayo, P., Mootha, V.~K., Mukherjee, S., Ebert, B.~L.,
  Gillette, M.~A., Paulovich, A., Pomeroy, S.~L., Golub, T.~R., Lander, E.~S.
  and Mesirov, J.~P. (2005) {Gene set enrichment analysis: a knowledge-based
  approach for interpreting genome-wide expression profiles.}
\newblock \emph{Proceedings of the National Academy of Sciences of the United
  States of America}, \textbf{102}, 15545--15550.

\bibitem[{Tibshirani(1996)}]{tibshirani96regression}
Tibshirani, R. (1996) Regression shrinkage and selection via the {L}asso.
\newblock \emph{Journal of the Royal Statistical Society, Series B},
  \textbf{58}, 267--288.

\bibitem[{Vuong(1989)}]{vuong1989}
Vuong, Q.~H. (1989) Likelihood ratio tests for model selection and non-nested
  hypotheses.
\newblock \emph{Econometrica}, \textbf{57}, 307--333.

\bibitem[{Wasserman and Roeder(2009)}]{wasserman2009}
Wasserman, L. and Roeder, K. (2009) High-dimensional variable selection.
\newblock \emph{Annals of {S}tatistics}, \textbf{37}, 2178--2201.

\bibitem[{Yuan and Lin(2006)}]{yuan06model}
Yuan, M. and Lin, Y. (2006) Model selection and estimation in regression with
  grouped variables.
\newblock \emph{Journal of the Royal Statistical Society, Series B},
  \textbf{68}, 49--67.

\bibitem[{Zou and Hastie(2005)}]{zou05regularization}
Zou, H. and Hastie, T. (2005) Regularization and variable selection via the
  elastic net.
\newblock \emph{Journal of the Royal Statistical Society, Series B},
  \textbf{67}, 301--320.

\end{thebibliography}
\end{document}